\documentclass[aip,jcp,reprint,showpacs,amsmath,amssymb]{revtex4-1}
\pdfoutput=1 

\usepackage{tikz}
\usepackage{graphicx}
\usepackage{dcolumn}
\usepackage{bm}
\usepackage{upgreek}
\usepackage{array}
\usepackage{booktabs}
\usepackage{microtype}
\usepackage{cmap}
\usepackage{geometry}
\usepackage{siunitx}

\newcommand{\casea}{{\bf A}}
\newcommand{\caseb}{{\bf B}}
\newcommand{\casec}{{\bf C}}
\newcommand{\cased}{{\bf D}}

\newcommand{\vect}[1]{\boldsymbol{#1}}

 \DeclareSIUnit{\cal}{cal}
 \DeclareSIUnit\Molar{\textsc{m}}

\usepackage{hyperref}
\hypersetup{colorlinks, citecolor={blue}, linkcolor={red}, urlcolor={violet}, pdftitle={Characterizing the bending and flexibility induced by bulges in DNA duplexes}, pdfauthor={J. S. Schreck, T. E. Ouldridge, F. Romano, A. A. Louis, J. P. K. Doye}, pdfdisplaydoctitle}

\begin{document}

\author{John S. Schreck}
\email{john.schreck@chem.ox.ac.uk}
\affiliation{Physical and Theoretical Chemistry Laboratory, Department of Chemistry, University of Oxford, South Parks Road, Oxford OX1 3QZ, United Kingdom}

\author{Thomas E. Ouldridge}
\affiliation{Rudolph Peierls Centre for Theoretical Physics, University of Oxford, 1 Keble Road, Oxford OX1 3NP, United Kingdom}

\author{Flavio Romano}
\affiliation{Physical and Theoretical Chemistry Laboratory, Department of Chemistry, University of Oxford, South Parks Road, Oxford OX1 3QZ, United Kingdom}

\author{Ard A. Louis}
\affiliation{Rudolph Peierls Centre for Theoretical Physics,  University of Oxford, 1 Keble Road, Oxford OX1 3NP, United Kingdom}

\author{Jonathan P. K. Doye}
\email{jonathan.doye@chem.ox.ac.uk}
\affiliation{Physical and Theoretical Chemistry Laboratory, Department of Chemistry, University of Oxford, South Parks Road, Oxford OX1 3QZ, United Kingdom}

\title{Characterizing the bending and flexibility induced by bulges in DNA duplexes}
\begin{abstract}

Advances in DNA nanotechnology have stimulated the search for simple motifs that can be used to control the properties of DNA nanostructures. One such motif, which has been used extensively in structures such as polyhedral cages, two-dimensional arrays, and ribbons, is a bulged duplex, that is two helical segments that connect at a bulge loop. We use a coarse-grained model of DNA to characterize such bulged duplexes. We find that this motif can adopt structures belonging to two main classes: one where the stacking of the helices at the center of the system is preserved, the geometry is roughly straight and the bulge is on one side of the duplex, and the other where the stacking at the center is broken, thus allowing this junction to act as a hinge and increasing flexibility. Small loops favor states where stacking at the center of the duplex is preserved, with loop bases either flipped out or incorporated into the duplex. Duplexes with longer loops show more of a tendency to unstack at the bulge and adopt an open structure. The unstacking probability, however, is highest for loops of intermediate lengths, when the rigidity of single-stranded DNA is significant and the loop resists compression. The properties of this basic structural motif clearly correlate with the structural behavior of certain nano-scale objects, where the enhanced flexibility associated with larger bulges has been used to tune the self-assembly product as well as the detailed geometry of the resulting nanostructures. 

\end{abstract}


\maketitle

\section{ Introduction }
DNA is one of the most important molecules in biology. By virtue of the specificity of Watson-Crick base-pairing,\cite{watson} DNA is highly programmable, and beginning with the work of Nadrian Seeman in the 1980s,\cite{seeman1982nucleic} it has been identified as a major player in nanotechnology, with many structures and devices already realized.\cite{bath2007dna, linko2013enabled} For example, by starting with appropriate oligonucleotides, DNA can be made to self-assemble into many structures with high yields by cooling solutions from high temperatures. The target structure is usually designed to be the global free energy minimum by virtue of containing the largest number of base pairs.  In many examples, the structures are made up of a few basic components, including double helical sections and various types of junctions where the double helices meet. Examples of these nanostructures include DNA polyhedra, such as tetrahedra,\cite{goodman2005rapid,he2008hierarchical,iinuma2014polyhedra} cubes,\cite{zhang2009symmetry, iinuma2014polyhedra} octahedra,\cite{shih20041, he2010chirality} icosahedra,\cite{zhang2008conformational} dodecahedra,\cite{he2008hierarchical} buckyballs,\cite{he2008hierarchical} and nanoprisms.\cite{zhang2012controlling,iinuma2014polyhedra} Alternatively, the ``DNA origami'' or ``DNA brick''  techniques allow for the construction of an enormous range of shapes built from closely-packed helices.\cite{rothemund2006folding, douglas2009self,ke2012three}  Making many of these structures is only possible because of DNA's physiochemical properties, which can be manipulated and therefore controlled.  Thus, understanding the biophysical and chemical properties of DNA is of vital importance for realizing future nanodevices and nanostructures.  

	\begin{figure}[t]
	\begin{center}
	\vspace{0.6 cm}
	\includegraphics[width =\columnwidth ]{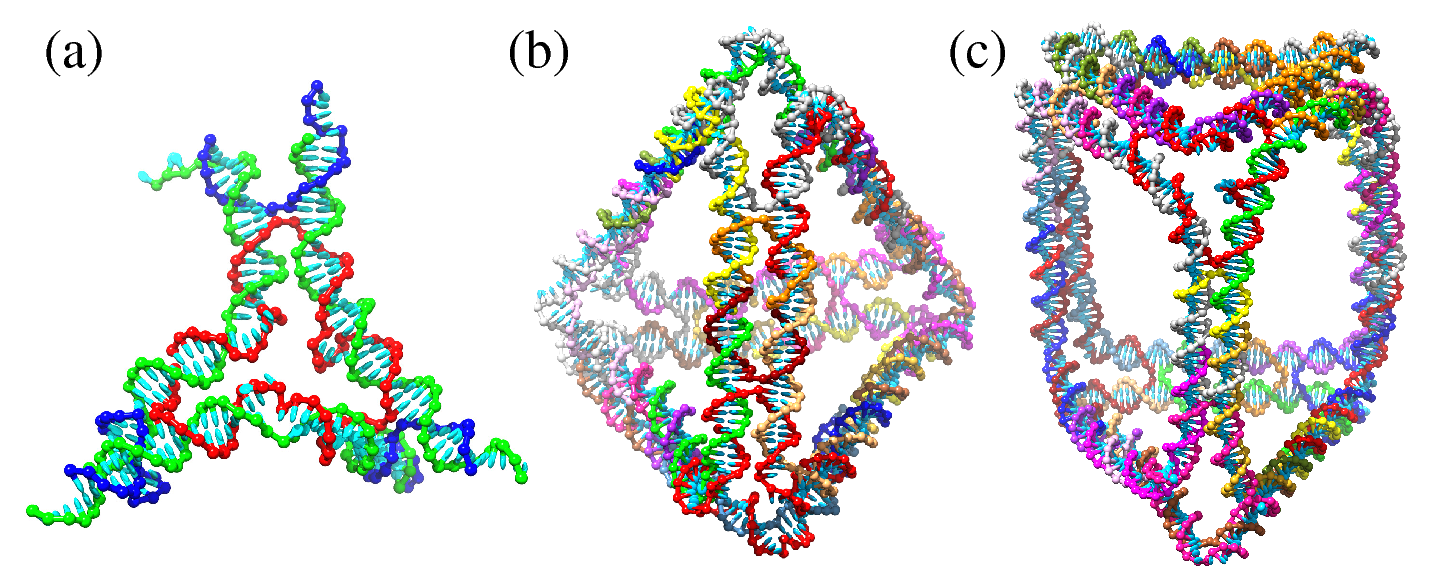}
	\caption{Representations of nanostructures containing bulges of varying size generated with oxDNA. (a) A 3-arm star tile containing three bulge regions at the center of the tile (red strand), where each bulge contains five nucleotides. (b) A DNA tetrahedron that is formed from four of the 3-arm star tiles shown in (a). (c) A DNA nanoprism made from six 3-arm tiles, whose chirality can be controlled by varying the sizes of bulges in constituent tiles.}
	\label{chiral_examples}
	\end{center}
	\end{figure}
A common motif used in many nanostructures is a 2-way junction in which two double helices are connected by a non-complementary bulge loop of varying size.\cite{riordan1992kinking, lilley1995kinking, lilley2000structures, bindewald2008rnajunction, cruz2009dynamic, bailor2010topology} Adjusting the bulge size in these 2-way junctions provides a way to control their flexibility. For example, a 3-arm DNA ``star-tile'' design,\cite{zhang2009dna} illustrated in Fig.~\ref{chiral_examples}(a), contains a combination of 2- and 4-way structural elements linked together by double-helical sections.\cite{ortiz1999crystal, lilley2009four, laing2009analysis, han2013dna} Where the arms meet at the center of the tile, the inner strand  (red in Fig.~\ref{chiral_examples}(a)) contains bulge loops in order to facilitate the bending of the arms with respect to one another. When a solution of identical star tiles is prepared and cooled, the tiles can self-assemble by linking together at the single-stranded `sticky' ends at the end of each arm. The assembly product is largely controlled by the bulge size.\cite{zhang2012controlling} For example, 3-arm tiles containing five nucleotides in each bulge region assemble into tetrahedra (illustrated in Fig.~\ref{chiral_examples}(b)), while the reduced flexibility of 3-arm tiles with three nucleotides in the bulge regions leads to the formation of dodecahedra or buckyballs, depending on the tile concentration.\cite{he2008hierarchical}  Another example is the DNA nanoprism illustrated in Fig.~\ref{chiral_examples}(c), where the sizes of the bulges between the different arms can be used to control the detailed geometry of the prism, in particular the relative twist of the top and bottom faces.\cite{zhang2012controlling} These examples illustrate that small changes in sequence design can clearly have a strong influence on the final product as well as the product structure. Even though bulged duplex structures are already widely used in nanodesigns, their effects on the structure and flexibility have not been thoroughly studied and the underlying principles have not been explored.  Meanwhile, strong bending of DNA duplexes in the absence of bulges continues to be an area of considerable interest.\cite{vologodskii2013strong}  

Early experimental studies showed that both DNA and RNA duplexes are subject to bending in the presence of bulges and internal loops, with bulges containing $\sim$5-7 nucleotides inducing larger bends than smaller bulges.\cite{hsieh1989deletions, riordan1992kinking, gohlke1994kinking, lilley1995nomenclature, zacharias1995bulge, zacharias1996influence, lilley2000structures} Subsequent studies showed that the induced bend is not rigid and the junction can adopt a variety of conformations with similar bend angles, which signifies an increase in flexibility caused by the presence of a bulge.\cite{al2002concerted,kim2002mg2} Following initial computational studies of A-form RNA duplexes in two-way junctions,\cite{chu2009conformational} Bailor \textit{et al.~}\cite{bailor2010topology, bailor2011topological, bailor20113d, mustoe2012new, mustoe2014coarse} have recently applied three inter-helical (Euler) angles that describe the bending and twisting flexibility at the two-way junctions: one angle quantifies the bend at the bulge, and the other two describe the relative twist of each of the duplex arms meeting at the bulge. Their studies of the Protein Data Base (PDB) have shown that the geometric secondary structure of 2-way junctions restricts the overall 3D orientation of helices in RNA,\cite{bailor2010topology} where these constraints arise from the steric and connectivity constraints imposed by the junctions.\cite{mustoe2012new, mustoe2014coarse}

The Euler angles provide a convenient and consistent way to characterize large numbers of configurations of 2-way junctions, which can be used to elucidate the level of influence the constraints imposed by the secondary structure have on the global conformation of DNA and RNA structures. In addition, several experimental groups have carried out similar analyses of two-way junctions in DNA,\cite{gohlke1994kinking, dornberger1999solution, stuhmeier2000fluorescence} where, for example, in recent work by Wozniak \textit{et al.},\cite{wozniak2008single} the three Euler angles for bending and twisting have been inferred from FRET experiments for bulged DNA duplexes with varying bulge size.  However, even with these recent experimental advances, much of the literature discussing experimental and theoretical investigations of two-way junctions has to date focused primarily on RNA. 

In this article, we use oxDNA, a coarse-grained model at the nucleotide level, to study in detail the structure and thermodynamics of B-DNA double helices containing a bulged loop, a motif that is henceforth referred to as a bulged duplex. The results of our simulations are compared with recent experiments in which the bend and twist angles of bulged duplex systems have been measured.\cite{gohlke1994kinking, dornberger1999solution, stuhmeier2000fluorescence, wozniak2008single} The degree of coarse-graining in oxDNA allows us to study not just the equilibrium structure, but also the flexibility of these junctions.

The oxDNA model has been highly successful at reproducing structural, mechanical, and thermodynamic properties for single- and double-stranded DNA.\cite{ouldridge2011structural}  Moreover, applications to study the fundamental biophysics of DNA, including the kinetics of hybridization,\cite{ouldridge2013dna, schreck2014dna} toehold-mediated strand displacement,\cite{srinivas2013biophysics} the response to mechanical stress such as the over-stretching transition of dsDNA under tension,\cite{romano2013coarse} the formation of cruciform structures under negative twist,\cite{matek2012dna} and the role of topology in the formation of kissing hairpin complexes,\cite{romano2012effect} have confirmed the robustness of the model. Furthermore, the model has proved useful in providing physical insight into the action of DNA nanodevices, such as nanotweezers,\cite{ouldridge2010dna} and walkers,\cite{ouldridge2013optimizing,vsulc2012simulating} and is starting to be applied to characterize large DNA nanostructures.\cite{doye2013coarse} oxDNA captures the relative flexibility of single strands, which can adopt a variety of helical and non-helical structures due to a strand's ability to stack and unstack,\cite{ouldridge2011structural,vsulc2012sequence} and the comparatively stiff duplexes. Because oxDNA simultaneously captures these thermodynamic and geometric effects for both single and double strands, it is well-suited to studying how the interplay of such fundamental factors shapes the overall behavior of bulges.
 
\section{Model and Methods} 
\label{sec:models}

\subsection{oxDNA model}

	\begin{figure}
	\begin{center}
	\vspace{0.6 cm}
	\includegraphics[width = 245 pt ]{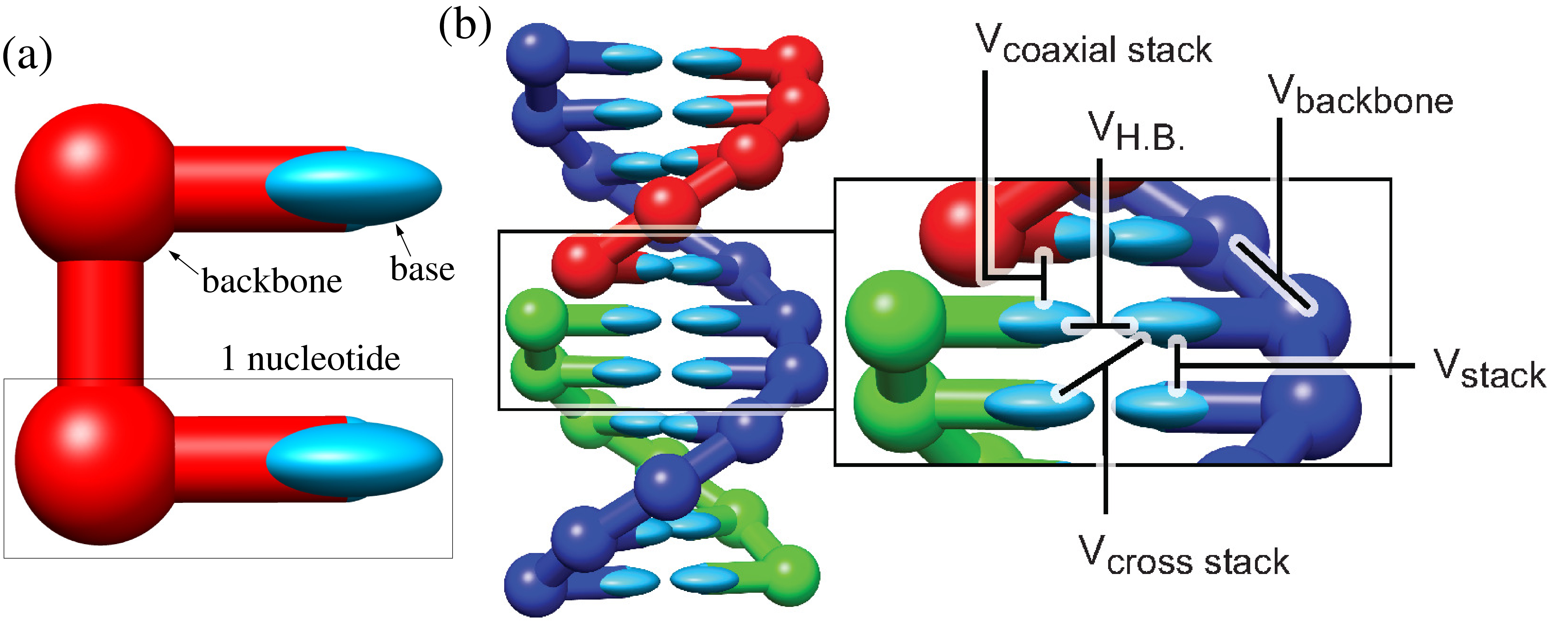}
	\caption{Simplified representation of (a) the rigid nucleotides that are the basic unit in oxDNA, and (b) an 11 base-pair double helix that illustrates the various interactions in the oxDNA model. Reproduced from Ref.~\onlinecite{doye2013coarse} with permission from the PCCP Owner Societies. }
	\label{oxDNA_model}
	\end{center}
	\end{figure}

In oxDNA, a single strand of DNA is modeled as a chain of rigid nucleotides, where each nucleotide contains one interaction site for the backbone and two more interaction sites associated with the stacking, coaxial stacking, and cross-stacking interactions.  The interactions in the model are illustrated in Fig.~\ref{oxDNA_model}. Base-pairing interactions contribute to the overall potential energy only when bases obey Watson-Crick specificity, e.g. A-T or G-C pairs. In addition to base pairing and stacking, there are interactions associated with backbone connectivity and excluded volume.  A detailed description of each interaction can be found in Ref.~\onlinecite{ouldridge2011structural}. Simulation codes for oxDNA are publicly available from the oxDNA website.\cite{oxdna_website}

The model does have have some simplifications that are important for the current study. First, the model was fit to experiments using a salt concentration of [Na$^+$] = \SI{0.5}{\Molar}, where the electrostatic interactions are strongly screened. Even though some studies have discussed the effect of salt concentration on the overall flexibility of bulged duplexes, the concentration [Na$^+$] = \SI{0.5}{\Molar} is in the high-salt regime that is relevant to most DNA nanotechnology experiments, and which is our primary interest. Secondly, the oxDNA double-helix is symmetrical with major and minor grooves being of the same size. While this may be important for certain motifs, we do not expect this approximation to change the dominant physics underlying bulge behavior, or the generic trends as a function of bulge size. Additionally, we choose to use the ``average-base'' parametrization of oxDNA, in which the hydrogen bonding associated with base pairing and the stacking interactions have the same strength independent of the bases involved, rather than its sequence-dependent parametrization.\cite{vsulc2012sequence} This parametrization is advantageous for studying general properties of DNA unmodulated by sequence-dependent effects.

\subsection{DNA bulged duplex systems studied}
\label{systems}
In our simulations we first consider a system where two DNA strands associate to form a bulged duplex, 
\begin{itemize}
\item 5$^\prime$ -- \textcolor{red}{CTA GCC TTGC} (T)$_M$ \textcolor{green}{GGAT GCT ACC} -- 3$^\prime$,
\item 5$^\prime$ -- \textcolor{green}{GGT AGC ATCC} \textcolor{red}{GCAA GGC TAG} -- 3$^\prime$,
\end{itemize}
where each duplex arm flanking either side of the bulge contains 10 hybridized base pairs (red regions are complementary to each other, as are green regions), and the bulge region contains $M$ consecutive thymine (T) bases.  This structure is very similar to the bulged duplexes found in the star tiles discussed above,\cite{he2008hierarchical} which contain 10 and 11 nucleotides in the arms, respectively. To investigate the effects of the size of the bulge loop on the structural properties we will consider $M$ ranging from zero to fifteen in our simulations.  Additionally, the simulation temperature was set at \SI{23}{\degreeCelsius}, which is near the temperature (\SI{25}{\degreeCelsius}) where the DNA nanoprism was found at high yield in experiments.\cite{zhang2012controlling}

In addition, we also consider C and Z-tiles,\cite{maoztile}  which are duplexes containing two bulge loops.  The following strand can be used to form a Z-tile, 
\begin{list}{\labelitemi}{\leftmargin=1em}
\item 5$^\prime$ -- \textcolor{red}{CTAACC}\textcolor{black}{ACTGG}\textcolor{green}{TGTCCGGACA}\textcolor{red}{GGTTAG}\textcolor{black}{CCAGT} -- 3$^\prime$,
\end{list}
where the red regions are complementary, as are the black regions.  The central green region is a palindrome. Two of these strands can hybridize through the red and green regions, where the final structure contains one green duplex section, two red duplex sections, and 4 single-stranded black sections (two bulges and two sticky ends). Note that the green duplex section is about one turn in length.  In general, the number of helical turns in the green duplex section determines the motif structure, where an even or odd number of half-turns yield Z- or C-tiles, respectively. The red sticky-ends are specifically designed to be complementary to the nucleotides in the bulge loop. Under certain conditions, the sticky ends may bind with loops from nearby tiles and can form a T-junction, that is, a region in which 3 helical arms meet at a bulge that resembles a T shape, which facilitates the self-assembly of the Z or C tiles into 1D or 2D nanostructures, respectively.  The Z-tile is further discussed in Section~\ref{sec:ztile}. The simulation temperature for Z-tiles was set at \SI{22}{\degreeCelsius}, the temperature at which the assembled tiles produced 1D nanostructures in high yield.\cite{maoztile}

\subsection{Simulation details}
To calculate free energies of structures and investigate flexibility, bulged duplexes for several bulge sizes $M$ are simulated by employing virtual-move Monte Carlo (VMMC), as introduced by Whitelam and co-workers.\cite{whitelam2009role} We use the variant introduced in the appendix of Ref.~\onlinecite{whitelam2009role}. VMMC is a cluster-move algorithm that efficiently samples from the canonical ensemble for systems of strongly-interacting particles, which we found particularly useful for DNA.\cite{doye2013coarse} 

The relative free energy of the bulged duplexes was sampled as a function of two order parameters: (1) the total number of base pairs in the system, and (2) the end-to-end distance, $R_{\text{ee}}$, defined as the distance between the center of mass of the bases at the 5$^\prime$ and 3$^\prime$ ends of the strand not containing the bulge.  Two bases are considered to be paired when the hydrogen bonding interaction between them is 0.093 times its well-depth (\SI{0.596}{\kilo\cal\per\mol} at \SI{23}{\degreeCelsius}). Small variations in this cutoff do not significantly affect the results. OxDNA predicts that the base pairs at the ends of double helices may temporarily break, a process termed fraying, even for systems well below their melting temperatures. Since this phenomenon may occur at either terminal end of a bulged duplex, or in the middle near the bulge loop, the effects that fraying may have on bending can be monitored by following changes in the number of base pairs. The end-to-end distance, $R_{\text{ee}}$, is useful for monitoring how much the system bends, with strong bending occurring when $R_{\text{ee}}$ is small compared to the length of a relaxed duplex. 

	\begin{figure*}
	\begin{center}
	\vspace{0.6 cm}
	\includegraphics[width = 490 pt ]{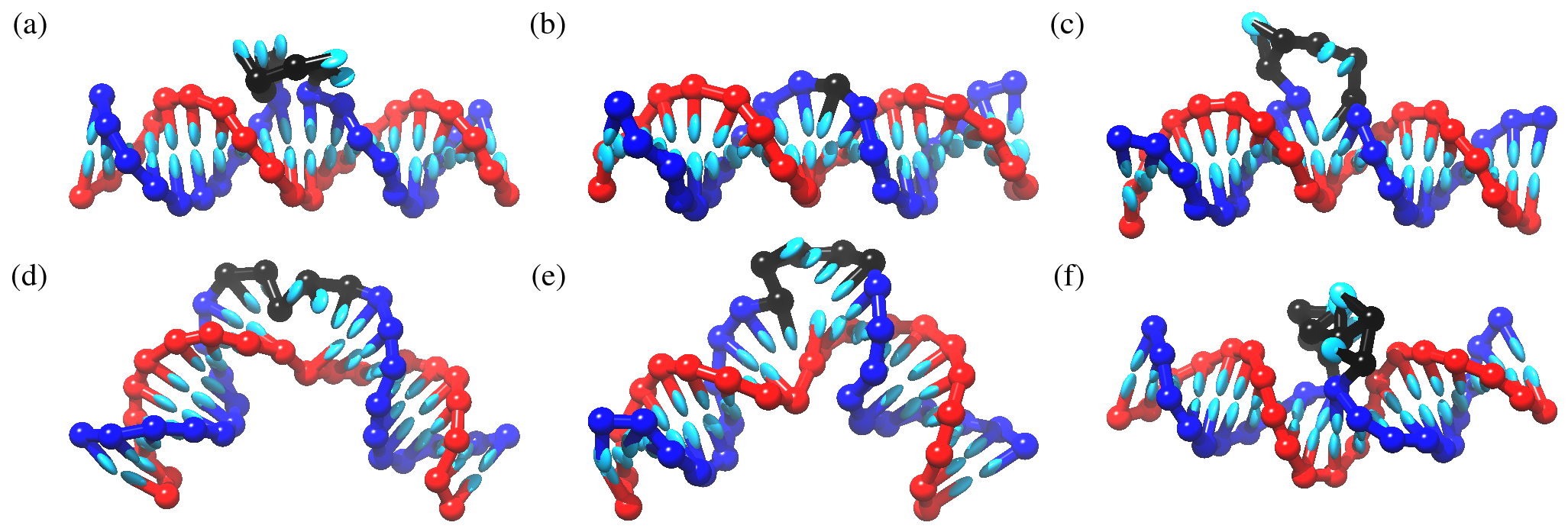}
	\caption{Configurations of bulged duplex: (a) a configuration classified as \casea{} in which bases stack across and opposite to the bulge ($M=5$); (b) a configuration classified as \caseb{} where a bulge loop of size $M=1$ is inserted into the duplex; (c) a configuration ($M=5$) classified as \casec{} where the stacking across the bulge is broken, but not opposite to the bulge, and one base from the loop is inserted inside the duplex; (d) a configuration ($M=5$) classified as \cased{}, where the stacking across and opposite to the bulge are broken. Both stem-loop stacks are intact; (e) a configuration ($M=5$) that has its helix interrupted due to the fraying of a base pair near the bulge; (f) a configuration ($M=5$) that is bent into the bulge. For clarity, all backbone elements in the bulge loop are colored black.}
	\label{stacking_cases}
	\end{center}
	\end{figure*}

 \subsection{Stacking and bulge classes}

In our simulations, the bulged duplex can adopt a variety of conformations for different bulge sizes, some of which are illustrated in Fig.~\ref{stacking_cases}, where the stacking at the center of the duplex system may or may not be interrupted. For example, in Fig~\ref{stacking_cases}(a) the stacking interaction between the two bases that flank either side of the bulge, which are not neighbors along the sequence, is intact. The stacking interaction between the two bases directly opposite to the bulge, which are neighbors, is also intact and the system is mainly straight. In the model, the stacking interaction between the bases flanking the bulge is referred to as a coaxial stack to distinguish it from the stacking interaction between two neighboring bases, such as the stack opposite to the bulge. Fig~\ref{stacking_cases}(b) illustrates a configuration where a 1-base bulge has become inserted into the helix and stacks with its neighboring bases that are adjacent to the bulge. The stack opposite the bulge is intact and the system is mainly straight. In Fig~\ref{stacking_cases}(c), a configuration is shown where the coaxial stack is broken and a base from the loop is inserted into the helix while the rest of the bulge bases are outside of the helix. The stack opposite from the bulge is intact, and only marginal bending of the system away from the bulge is observed. In Fig~\ref{stacking_cases}(d), both the coaxial stack and the stack opposite to the bulge are broken and the system is bent away from the bulge. Other illustrated states of the system include a configuration in which fraying disrupts the base pairs flanking the bulge as well as interrupting stacking along the helix (Fig~\ref{stacking_cases}(e)), and a configuration which bends towards the bulge (Fig~\ref{stacking_cases}(f)). Similar observations are made for the Z-tile at each of the two bulged regions. 

We see four main classes of conformations for the bulged duplex. The classifications are:
\begin{description}
\item[\textbf{A}] The bases in the loop disrupt the duplex as little as possible and are flipped out, which results in almost no bending.
\item[\textbf{B}] The bases from the bulge loop are inserted into helix while maintaining stacking opposite the bulge, resulting in some degree of static bending away from the bulge.
\item[\textbf{C}] A combination of \casea{}  and \caseb{} in which some bases are inserted into the helix, while others are flipped out.
\item[\textbf{D}] The stacks opposite and across from the bulge are broken, resulting in increased flexibility and a large static bend.
\end{description}  

\begin{figure}
	\begin{center}
	\vspace{0.6 cm}
	\includegraphics[width = \columnwidth]{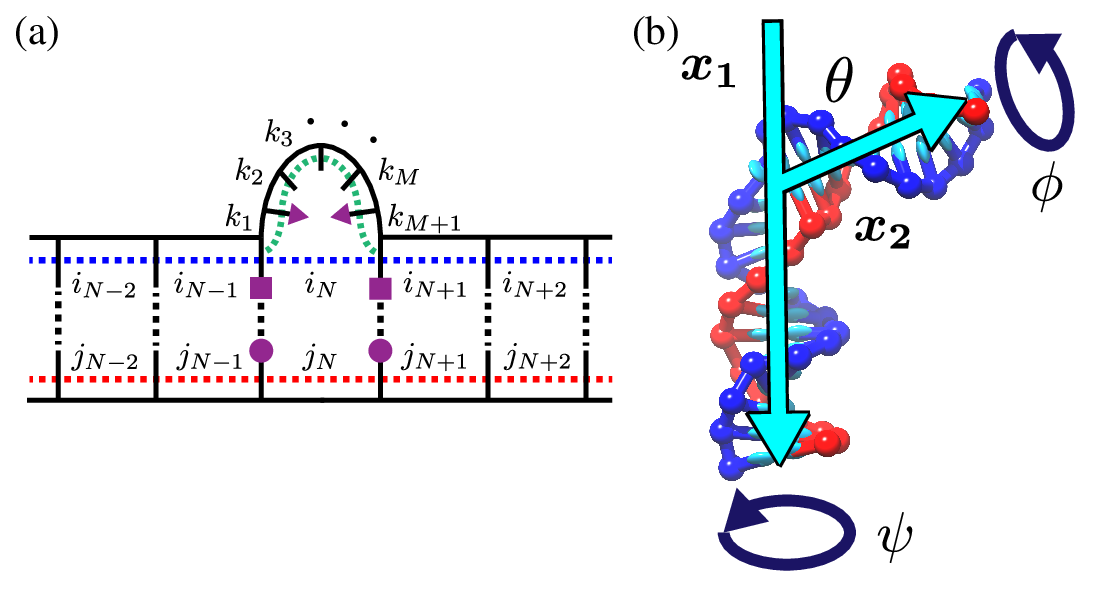}
	\caption{(a) Illustration of the interactions near the bulge. Short, solid black line segments represent bases, dashed blue lines represent stacking interactions between neighboring bases on the strand containing the bulge (top strand, bases next to the bulge are indicated by squares), and are labeled using the symbol $i$, while dashed red lines represent stacking interactions between neighboring bases on the opposite strand (bottom strand, bases opposite the bulge are indicated by circles) and are labeled with the symbol $j$. Dashed green lines that are indicated by the symbol $k$ represent the stacks that involve bases from the bulge loop. Dashed black lines represent H-bond interactions between bases. (b) The bend angle $\theta$ and twist angles $\phi$ and $\psi$ are illustrated along with vectors $\vect{x_{1}}$ and $\vect{x_{2}}$ that are used to define the angles.}
	\label{geometry}
	\end{center}
	\end{figure}

In order to apply these classifications, we focus on the stacking interactions between the pairs of bases near the bulge at the center of the system as well as base pairs that are adjacent to the bulge, as illustrated in Fig~\ref{geometry}. To determine whether a base stacks or coaxially stacks with another base, a lower bound for the stacking interaction between two bases is defined to be the same as the hydrogen bonding cutoff value of \SI{0.596}{\kilo\cal\per\mol} at \SI{23}{\degreeCelsius}. The choice of this value is somewhat arbitrary, however, small changes in this value do not change the results significantly. 

In Fig~\ref{geometry}(a), the state of the coaxial stacking interaction between bases across the bulge is denoted by the symbol $i_N$, while the stacking states of the two bases on either side of the bulge are denoted $i_{N-1}$ and $i_{N+1}$, respectively. Likewise, the state of the stack opposite the bulge is denoted $j_N$, while the state of the stacks between neighboring bases on the same strand on either side of the bulge are denoted $j_{N-1}$ and $j_{N+1}$, respectively. Finally, the state of the stacks between the bases with squares and the bases with triangles on either side of the bulge are denoted $k_{1}$ and $k_{M+1}$, respectively, and are referred to as ``stem-loop'' stacks, while the stacks between the bases in the loop are denoted $k_{2}, k_{3}\dots, k_{M}$. If the magnitude of the stacking interaction is greater than or equal to the lower bound, $i_{m}$ = 1, otherwise the stack is taken to be broken and $i_{m}$ = 0, and similarly for $j$ and $k$. 

If the stack opposite the bulge, $j_N$, is present, we take the two duplex arms to be ``stacked'' and group these configurations into a stacked set. Configurations where $j_N$ is disrupted are taken to be  ``unstacked'' and are grouped into an unstacked set. According to our classification scheme, a stacked configuration may fall into either \casea{}, \caseb{} or \casec{}, while an unstacked configuration is classified as \cased{}. The stacked configurations can be subdivided by taking into account the status of the stem-loop stacks that flank both sides of the bulge ($k_1$ and $k_{M+1}$), and the stacks between bases in the loop ($k_2, \dots, k_M$). A stacked configuration falls into \casea{} if both stacks at the stem-loop interfaces are broken (i.e.\ $k_1$ and $k_{M+1}$ are broken); \caseb{} if one or fewer stacking interactions are broken in the strand containing the bulge, counting all stacks from one stem-loop stack to the other, i.e.\ $k_1 \dots k_{M+1}$; and \casec{} if greater than one stack in the loop is broken but stacking is intact at one or both of the stem-loop stacks (more than one of $k_2, \dots, k_M$ broken and at least one of $k_1$ or $k_{M+1}$ intact). We note that configurations that have frayed base pairs at the junction can be difficult to classify using our scheme. In the supplemental materials we discuss several extensions to the classification scheme to properly deal with frayed configurations.\cite{supmat}

\subsection{Bend and twist angles of bulged duplexes}
\label{bendandtwist}
We define the bend and twist angles using a scheme similar to that used in Ref.~\onlinecite{bailor2010topology}. To measure the bend angle, $\theta$, for a given configuration we place unit vectors labeled $\vect{x_{1}}$ and $\vect{x_{2}}$ along the helical axes of each of the duplex arms flanking the bulge. These vectors are illustrated by blue arrows in Fig.~\ref{geometry}(a) and in more detail in supplemental Fig.~\ref{ztile_geometry_fig}(a). The duplex arms have the freedom to twist about the vectors $\vect{x_{1}}$ and $\vect{x_{2}}$ as characterized by the angles $\phi$ and $\psi$, respectively, as illustrated in Fig.~\ref{ztile_geometry_fig}(a).\cite{supmat} The vectors $\vect{x_{1}}$ and $\vect{x_{2}}$ are defined by finding the longest stretch of base pairs in each arm, and then drawing a line from the center-of-mass of the base pair at one end of the duplex arm to the center-of-mass of the base pair at the opposite end of the same duplex arm. For normalized $\vect{x_{1}}$ and $\vect{x_{2}}$, the bend angle $\theta$ is then calculated using
\begin{equation}
\vect{x_{1}} \cdot \vect{x_{2}} = -\cos \left( \theta \right),
\end{equation}
a definition that follows the convention for the bend angle in the recent literature.\cite{bailor2010topology} In the supplemental material,\cite{supmat} we develop a simple convention to determine whether the system is bent away from (\ang{0} $< \theta \le$ \ang{180}) or bent into the bulge (\ang{-180} $\le \theta <$ \ang{0}).

Lastly, we explicitly define the duplex twisting angles $\phi$ and $\psi$. Each angle can be calculated by first computing a vector that points from the base flanking the bulge to its complementary partner directly across from the bulge. The two vectors are referred to as $\vect{d_1}$ and $\vect{d_2}$ and point from square to circle in Fig.~\ref{geometry}(a) and are illustrated in Fig.~\ref{ztile_geometry_fig}(a).  The twist angles $\phi$ and $\psi$ can be calculated using
\begin{eqnarray}
\vect{d_1} \cdot \vect{z} &=&  \cos \left( \phi  \right), \\
\vect{d_2} \cdot \vect{z} &=&  \cos \left( \psi \right),
\end{eqnarray}
respectively, where $\vect{z} = \vect{x_{1}} \times \vect{x_{2}}$ is a vector normal to the plane of the bulged duplexes.  Similar to $\theta$, we develop a convention for determining when the angles $\phi$ and $\psi$ take on the values between \ang{0} $< \phi \le$ \ang{180} (\ang{0} $< \psi \le$ \ang{180}), and \ang{180} $< \phi \le$ \ang{360} (\ang{180} $< \psi \le$ \ang{360}). This is discussed in section S1 in the supplemental materials.  The relative twist between the duplex arms flanking the bulge is taken as $\phi-\psi$. For reference, in a relaxed duplex as represented by oxDNA, $\phi-\psi$ $\approx$ \ang{32}, i.e.\ the twist per base pair rise in a duplex. 

\section{Results}

\subsection{Bulged duplex systems}

\subsubsection{Stacking classes}

\begin{figure}
	\begin{center}
	\vspace{0.6 cm}
	\includegraphics[width = \columnwidth ]{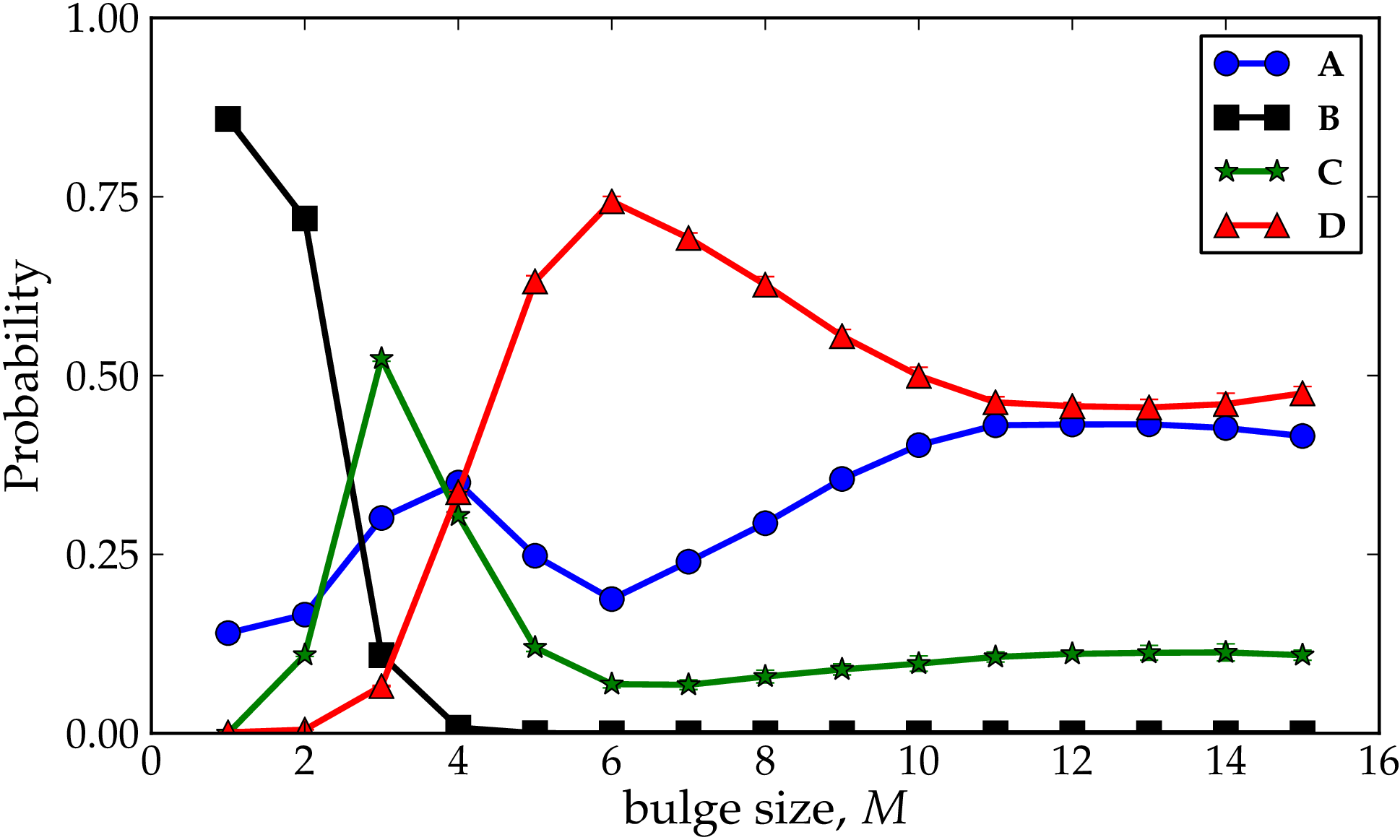}		
	\caption{The fraction of configurations found in structural classes \casea{}, \caseb{}, \casec{}, and \cased{} are plotted as a function of the bulge size $M$. In the plot the error bar on each point are smaller than the symbol size.}
	\label{fraction_plot}
	\end{center}
	\end{figure}

We first discuss how the balance between structural classes \casea{}, \caseb{}, \casec{}, and \cased{} changes as a function of the length of the bulge, $M$, in oxDNA. Each bulged duplex system with a given bulge length was simulated at least 10 times and configurations were collected until the standard error of the mean for the points computed in the average bend angle versus bulge size was less than 1\% of the the computed mean value. The results are shown in Fig.~\ref{fraction_plot}.

Systems with small bulge loops ($M=1$ and $M=2$) are dominated by class \caseb{}. In this scenario the bases in the bulge are inserted into the helix, as was illustrated in Fig.~\ref{stacking_cases}(b). Insertion is favored over flipping the bases outside of the helix because more stacking interactions are preserved. For $M=3$ it is no longer feasible to insert 3 bases into the duplex and maintain stacking opposite the bulge. Therefore, the probability of \caseb{} decreases dramatically, and instead configurations where one or two of the bases in the bulge are inserted into the duplex, while the others predominately flip out, become prevalent. Additionally, $M=3$ is the first case where unstacked states make a significant contribution. For $M>4$, class \cased{} (e.g. unstacked configurations) becomes the most probable class and of the stacked states, \casea{} becomes increasingly dominant over class \casec{} as $M$ increases. Also very apparent is that the probability of stacked states goes through a maximum at $M=6$ before plateauing off at $M\gtrsim11$ where cases \casea{} and \cased{} are roughly equally probable with some contributions still arising from case \casec{}. 

The most obvious trend is that although unstacked configurations are rare for small $M$, they become common at larger $M$. In stacked configurations, the endpoints of the bulge loop are strongly constrained; unstacked configurations offer greater freedom. Whether or not unstacking occurs depends on the relative benefit of this freedom compared to the cost of disrupting junction stacking. 

Loops with small $M$ do not benefit as much from the additional freedom of the unstacked ensemble for two reasons. Firstly, they are too short for all relative orientations of the duplex arms to be explored. As the loop gets longer, the duplex arms have more freedom and the unstacked state becomes more favorable. Secondly, steric penalties associated with the stacked configuration are smaller for small $M$. 

The bulged duplex systems that we study are somewhat similar to a system with one or two single-stranded dangling ends at a junction in a duplex, except that in our cases the strands are connected and form a bulge loop. Duplexes with such dangling ends can occur during toehold-mediated strand displacement when an invader strand is displacing an incumbent strand during branch migration, as described in Ref.~\onlinecite{srinivas2013biophysics}. In these systems, once the invading strand is bound to the toehold, there is a free-energy penalty to initiating displacement, even though the number of base pairs is unchanged, because it is unfavorable to have two single-stranded overhangs at the junction. The free-energy penalty arises from the overcrowding of nucleotides at the junction, and saturates once both overhangs have at least 3 or 4 nucleotides, because further bases are sufficiently far enough away from the junction that their contribution to overcrowding is minimal. 

Similar to the displacement system, for the bulged duplex to maintain a stacked state (i.e.\ one of cases \casea{}, \caseb{}, and \casec{}) the bulge loop must arrange itself to minimize steric clashing amongst the bulge bases and also minimize duplex disruption caused by inserted bulge bases which can compete with the coaxial stack. Alternatively, a bulged duplex may unstack and bend away from the bulge gap, freeing the bulge bases to spread out into space and decrease overcrowding. As with displacement intermediates, the benefit of spreading out is greater when there are more nucleotides at the junction, providing the second cause for the the increase in unstacking with increasing $M$. 

Several aspects of the data, however, are not explained by this analysis. Firstly, why does case \casec{} become less favorable compared to case \casea{} as loop length increases? Secondly, why does case \cased{} increase with respect to case \casea{}, and then subsequently decrease (which will subsequently give a non-monotonic variation of the bend angle as shown in the next section)? Both of these questions can be understood in terms of how changes in configuration at the junction are related to the typical physical length of the bulge. In order for the duplex arms to be stacked (in class \casea{}, \caseb{} or \casec{}), any bases not incorporated into the helix must be compressed so that the single-stranded loop region adopts a conformation with a short end-to-end distance. When the bulge loops are short ($M \le 4$), inserting one or two bases into the duplex helps to reduce the compression substantially: a 2-base loop is less constrained than a 3-base loop because it needs to be compressed less. As the loop gets larger, however, this difference becomes less substantial and so it is not as advantageous for a stacked duplex to incorporate bases. Thus case \casec{} is favorable for short loops, but less so for longer loops. 

Next we address the competition between cases \casea{} and \cased{}. Medium-sized loops (i.e.\ $5 \le M \le 10$) benefit more from unstacking than longer loops ($M \ge 11$) which resist compression less. This is because medium sized loops are only slightly longer than the persistence length of ssDNA, and unstacking of duplexes at the junction allows them to stretch out more easily. By contrast, for longer loops the relative cost of being bent is lower, and so they benefit a lot less from being unstacked at the junction. 

 This typical behavior of polymers can be seen, for example, in the end-to-end probability distribution of a worm-like chain.\cite{becker2010radial} The analytical formula for the probability density of the end-to-end distance in Ref.~\onlinecite{becker2010radial} clearly shows that a shorter polymer benefits a lot more from having its end-to-end extension increased by a fixed absolute distance when compared with a longer polymer, as illustrated in Fig.~\ref{wlc_plot}, where chains with contour lengths that are two and four times the persistence length are compared. Similar results can also be obtained for a freely-jointed chain. 

\begin{figure}
	\begin{center}
	\vspace{0.6 cm}
	\includegraphics[width = \columnwidth ]{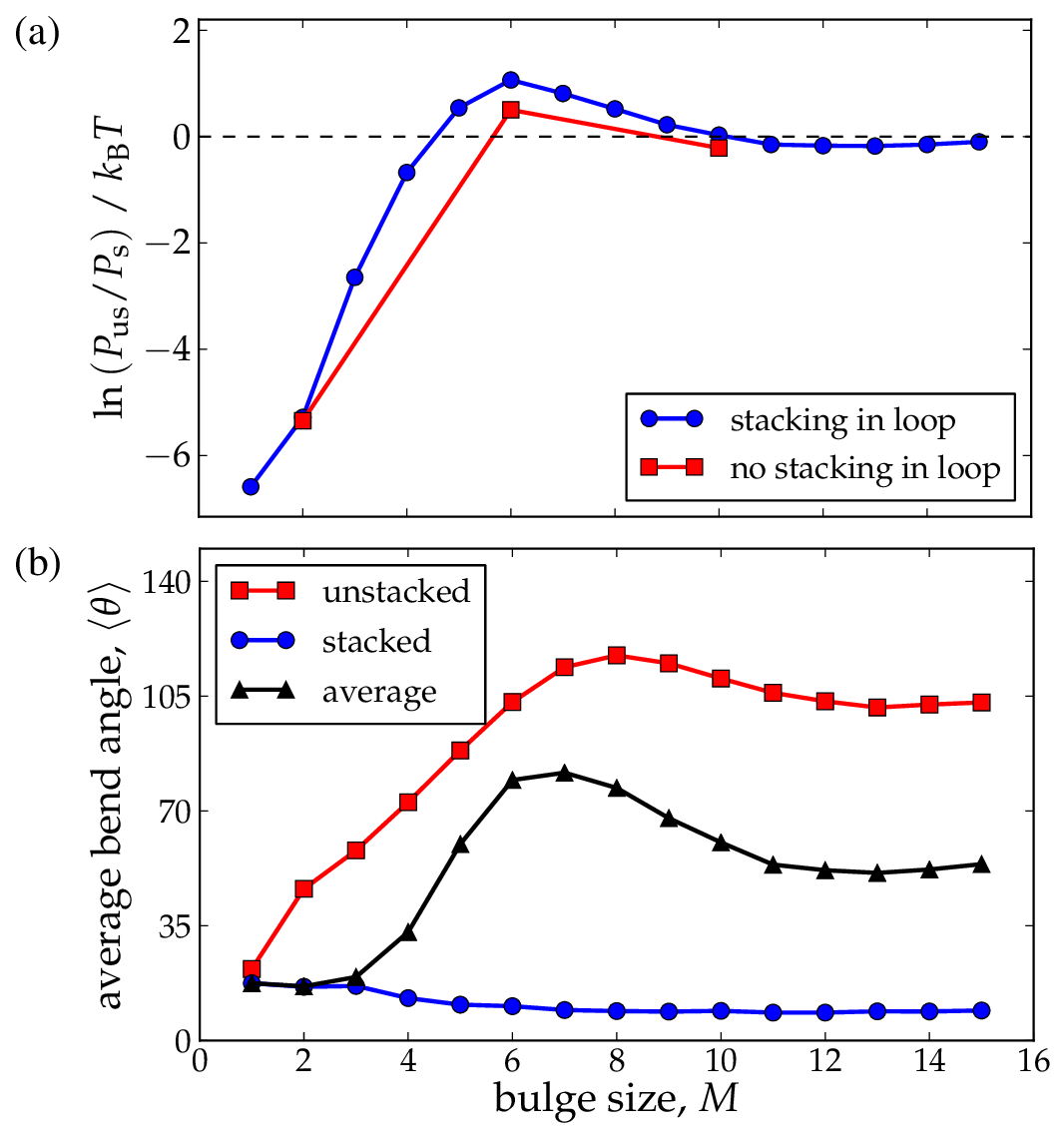}
	\caption{ (a) The relative difference in free energy between unstacked and stacked configurations. $P_{\text{us}}$ and $P_{\text{s}}$ are the probabilities that a bulged duplex is found in the unstacked and stacked states, respectively. (b) The average bend angle for stacked (red circles) and unstacked (blue squares) populations is plotted against bulge size $M$. The average bend angle for the complete set of configurations is also plotted (black triangles). In both figures the error bar on each data point are smaller than the symbol size.  }
	\label{ratio_M}
	\end{center}
	\end{figure}

We checked whether the non-monotonic behavior was peculiar to our model of ssDNA, or a generic polymer effect, by switching off nearest-neighbor stacking interactions between loop nucleotides (while maintaining them in the stem and at the interface of stem and loop). The results, shown in Fig.~\ref{ratio_M}(a), which plots the free-energy difference between the combined classifications \casea{}, \caseb{}, \casec{} and classification \cased{}, show that stacking between bases in the loop does not significantly change our results and therefore that our results are a robust consequence of generic polymer behavior.

\subsubsection{Effects of stacking on bending angle $\theta$}

	\begin{figure*}
	\begin{center}
	\vspace{0.6 cm}
	\includegraphics[width = 2 \columnwidth ]{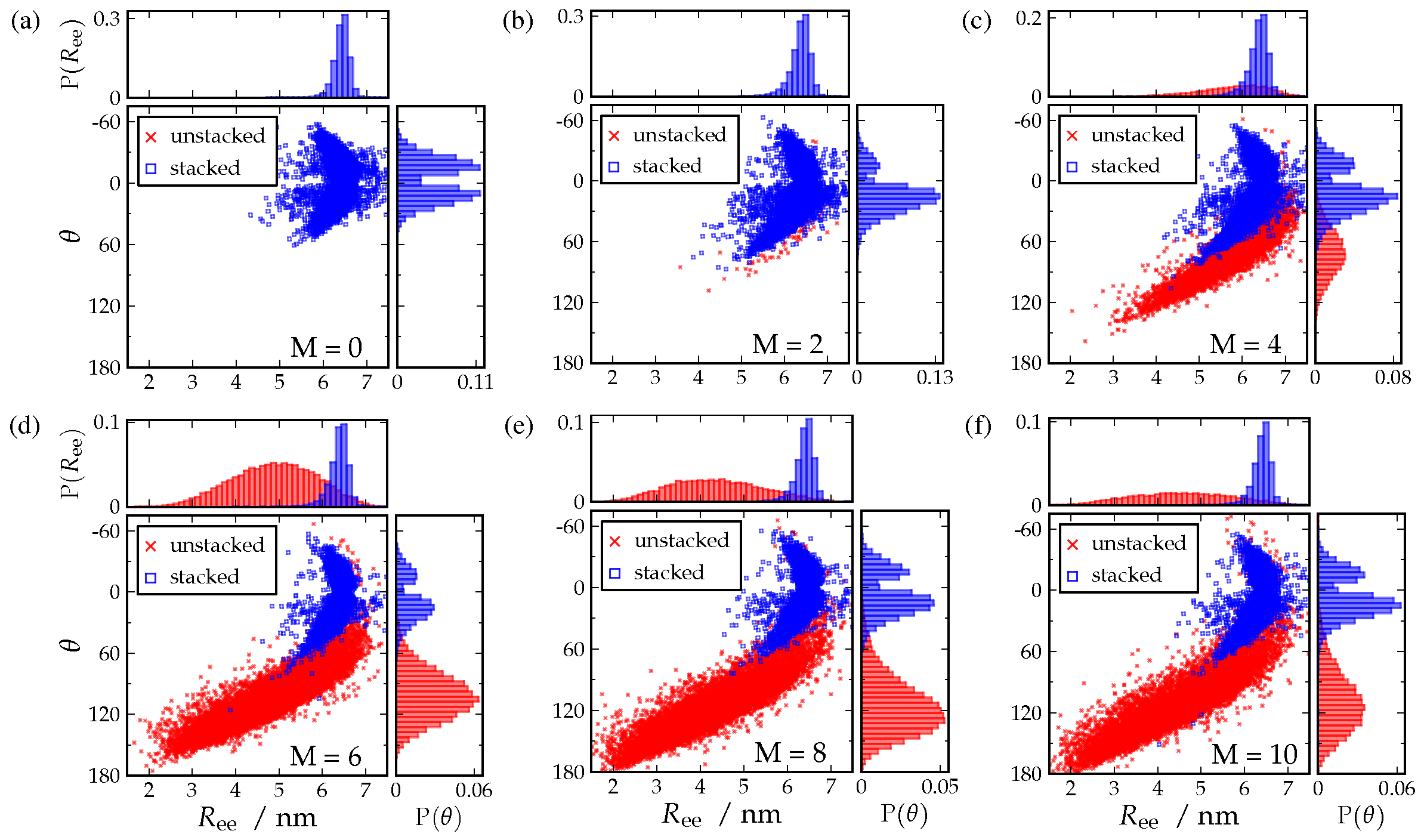}		
	\caption{In each figure there are three plots that illustrate the relationship between bend angle $\theta$ and end-to-end distance $R_{\text{ee}}$: a scatter plot in ($R_{\text{ee}}$,$\theta$) for sets of configurations generated in our simulations, the probability of a state occupying a particular value of $R_{\text{ee}}$ (top panel), and the probability of a state occupying some value of the bend angle, $\theta$ (right panel). Data is plotted for a selection of bulge sizes in the range $M=0$ to $M=10$. In the scatter plots, red crosses indicate unstacked states and blue squares represent stacked states. The probability plots retain the same color scheme as the scatter plots. Equivalent plots for other bulge sizes can be found in the supplemental materials.\cite{supmat}}
	\label{2d_plots}
	\end{center}
	\end{figure*}
Fig.~\ref{ratio_M}(b) shows the average bend angle $\langle \theta \rangle$ as a function of bulge size, as well as the bend angle for stacked (classes \casea{}, \caseb{}, and \casec{}) and unstacked states (class \cased{}). The average bend angle for a bulged duplex for small bulge sizes is small when most configurations are stacked, but it quickly grows with $M$ as unstacked configurations become more favorable. The average bend angle peaks at \ang{80} at $M=7$ before decreasing again and finally plateauing at about \ang{55} for $M\ge 11$, this behavior mainly reflecting the variation of the probability of being unstacked (Fig.~\ref{ratio_M}(a)).

The average bend angle for unstacked states increases monotonically from $M=1$ and reaches a maximum at $M=8$ where $\langle \theta \rangle =$ \ang{117}, but then starts to decrease as $M$ grows, eventually leveling off at $\langle \theta \rangle \approx$ \ang{105} for $M \ge 11$. The drop in the average bend angle before leveling off is also a signal of the compression effect. The configurations belonging to the stacked population start off slightly bent at \ang{20} due to the high probability that the coaxial stack is broken because a few bases have inserted into the helix, but the bending angle tends towards \ang{10} for longer loops because there is less incentive for the bases to insert into the helix. 

Another noteworthy feature is that the bend angle in stacked and unstacked junction configurations is essentially independent of the nearest-neighbor stacking between the bases in the loop (see Fig.~\ref{repulsive_effects}).\cite{supmat} The interactions at the stem-loop interface, however, are important. These stem-loop stacks influence the bending angle $\theta$ when configurations are classified as \cased{} (configurations in which duplex arms are unstacked at the junction). We demonstrated this by switching off the nearest-neighbor stacking interactions between bulge loop bases, the stem-loop stacks, the coaxial stack, and the stack opposite to the bulge. We compared this system to systems in which the stem-loop interactions were not removed. The results, presented in Fig.~\ref{repulsive_effects}, show that for unstacked configurations the stem-loop interactions cause the system to bend significantly more than if the stacks were not present.\cite{supmat} This is because such stem-loop interactions direct the loop bases to carry on in the direction of the stems, thus requiring a larger bend angle to avoid steric repulsion, as is clear from the configuration in Fig.~\ref{stacking_cases}(d). 

For a more detailed look at how changes in bulge size influence the flexibility of bulged duplexes, in Fig.~\ref{2d_plots} we plot the bend angle and end-to-end distance distributions for various $M$. For stacked configurations there is a clear asymmetry between bending into, and away from the bulge loop, with configurations that bend away from the bulge more favored because these configurations reduce steric clashing between the bulge loop and the duplex. This asymmetry is most pronounced for small $M$ where base insertion is prevalent. 

The separation of the stacked and unstacked states is clear from the scatter plots in Fig.~\ref{2d_plots}. As unstacked configurations have a slightly increased effective contour length relative to their stacked counterparts, the two distributions overlap more when projected onto the end-to-end distance, $R_{\text{ee}}$, than $\theta$. The range of bend angles that are available to unstacked states becomes wider as the bulge size increases, and underlies the increase of $\langle \theta \rangle$ with bulge size for $M = 1-8$. Comparing the bend angle distributions for $M=8$ with $M=10$ for the unstacked states, it is clear that the distribution for $M=10$ is less sharply peaked than the distribution for $M=8$, and underlies the slight decrease in the average bend angle $\langle \theta \rangle$ for the unstacked states when increasing the bulge size from $M=8$ to $M=10$. 

\subsubsection{Effects of stacking on duplex twist angle $\phi-\psi$}

	\begin{figure*}
	\begin{center}
	\vspace{0.6 cm}
	\includegraphics[width = 2\columnwidth]{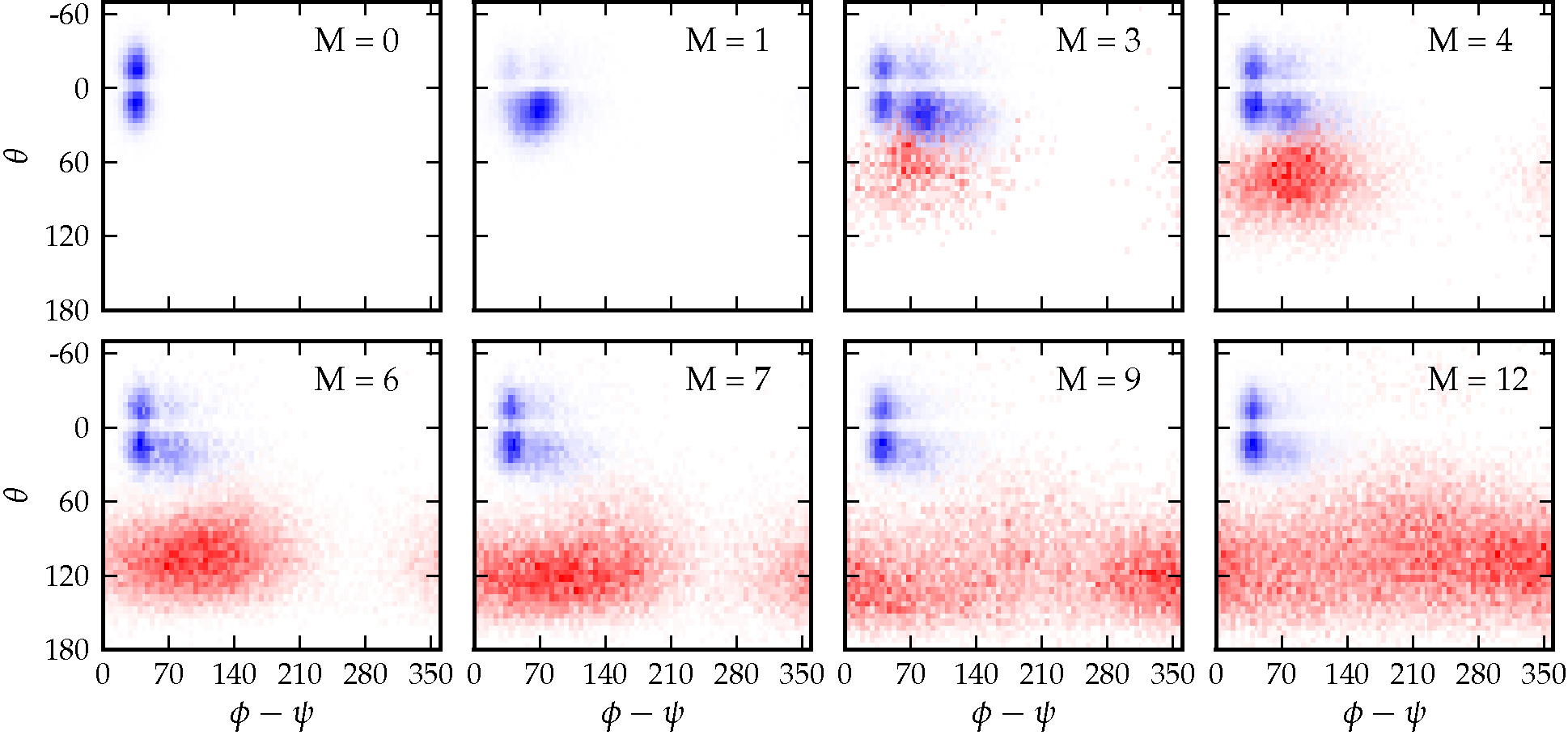}
	\caption{The probability that a configuration assumes some value of ($\theta$, $\phi$-$\psi$) is plotted for various bulge sizes $M$. Configurations were separated based on the state of the stacking between the double stranded regions, with the stacked set in blue and the unstacked set in red. The probability distributions for the stacked and unstacked sets were normalized to one separately to enhance clarity. Equivalent plots for other bulge sizes can be found in the supplemental materials.\cite{supmat}}
	\label{twist_plot}
	\end{center}
	\end{figure*}

To quantify the relative twisting of the duplex arms, $\phi-\psi$ is calculated for the same set of configurations.  In Fig.~\ref{twist_plot}, the probability that a configuration occupies a state with a given value of ($\theta$, $\phi-\psi$) is plotted as a 2D histogram. As with the analysis of the bend angle, $\theta$, we split the configurations into stacked and unstacked sets. A very pronounced (and expected) feature common in all bulged duplexes we considered is a stacked duplex population with a relative twist that is found to lie in a very narrow range, centered at approximately \ang{32}, and is due to configurations in class \casea{}.  

As illustrated in Fig.~\ref{twist_plot}, once a bulge is introduced into the duplex, the relative twist at the bulge maintains a signal at \ang{32} for all values of $M$ studied, however a second signal at approximately $\phi-\psi \approx$ \ang{70} appears for stacked configurations that are mainly bent away from the bulge. These configurations mostly correspond to the scenario where some number of the bases from the loop are inserted into the duplex at the bulge, as illustrated in Fig.~\ref{stacking_cases}(b) and (e). As we had shown earlier for all bulged duplexes, one, and sometimes two bases can be inserted into the duplex. Clearly, extra bases at the center of the duplex cause the bulged duplex to twist more at the bulge than stacked configurations belonging to class \casea{}. The spread for the relative twist angles in these states is also quite narrow and the effects due to base insertion is greatest for $M=1$ loops when the base in the bulge is much more likely to be found inserted rather than flipped outside of the duplex. Base insertion significantly persists up to $M=4$, then the effect decreases for $M > 4$ and plateaus. These plots nicely illustrate the fact that there remains a finite probability at large $M$ that a base from a longer loop can still become inserted into a duplex, as can be seen in Fig.~\ref{stacking_cases}(c).
 
Fig.~\ref{twist_plot} also shows some interesting features of the unstacked bent states. Initially, when the unstacked population first appears with a significant probability compared to the stacked state ($M=4$), the relative twist angles roughly fall in a circular distribution in the $\theta-(\phi-\psi)$-plane, centered around $\theta \approx$ \ang{70} and $\phi-\psi \approx$ \ang{100}.  However, deviations from the center are modest and are increasingly less probable due to the constraint of the short loop. There is also some overlap between stacked and unstacked populations for $M=4$, which corresponds to those configurations where the coaxial stack is broken, but not the stack opposite the bulge, and those configurations where both stacks are broken, but the bending angle is still constrained by the geometry of the loop, respectively. Upon increasing the bulge size further from $M=6$, the circular distribution elongates along the $\phi-\psi$ axis, because the longer loop allows the bulged duplex arms considerably more freedom to twist relative to each other. Continuing this trend, once the bulge size increases up to $M \ge 11$, the duplex arms can be oriented at almost any relative twist angle with relative ease.

\subsection{Z-tile structure}
\label{sec:ztile}

    \begin{figure*}
	\begin{center}
	\vspace{0.6 cm}
	\includegraphics[width = 2\columnwidth ]{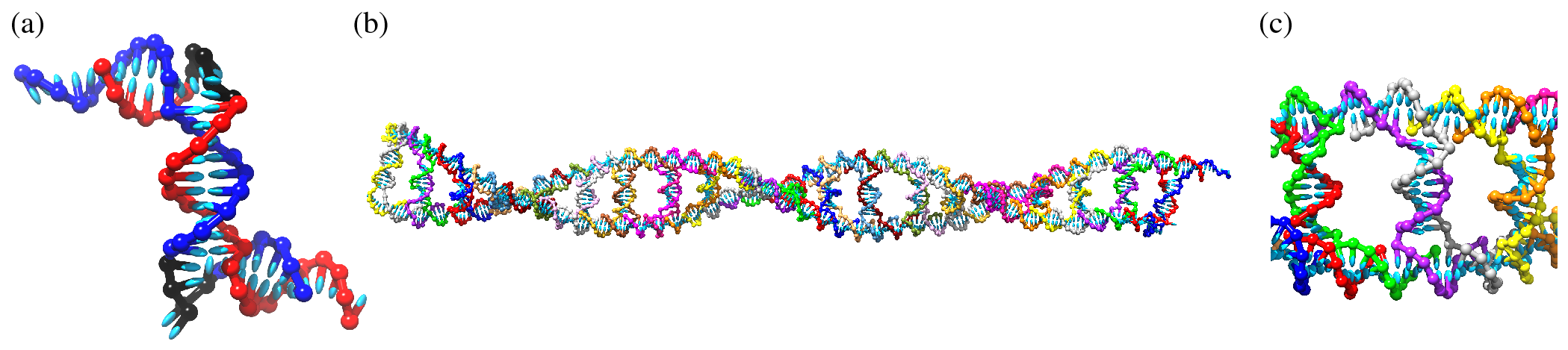}
	\caption{ (a) Illustration of a Z-tile where the structure is bent at both bulges. For clarity, the two bulge regions are colored black. (b) An assembly of twenty Z-tiles that are linked together at T-junctions. (c) Assembled structure of T-junctions.}
	\label{ztile}
	\end{center}
	\end{figure*}

    \begin{figure}
	\begin{center}
	\vspace{0.6 cm}
	\includegraphics[width = \columnwidth ]{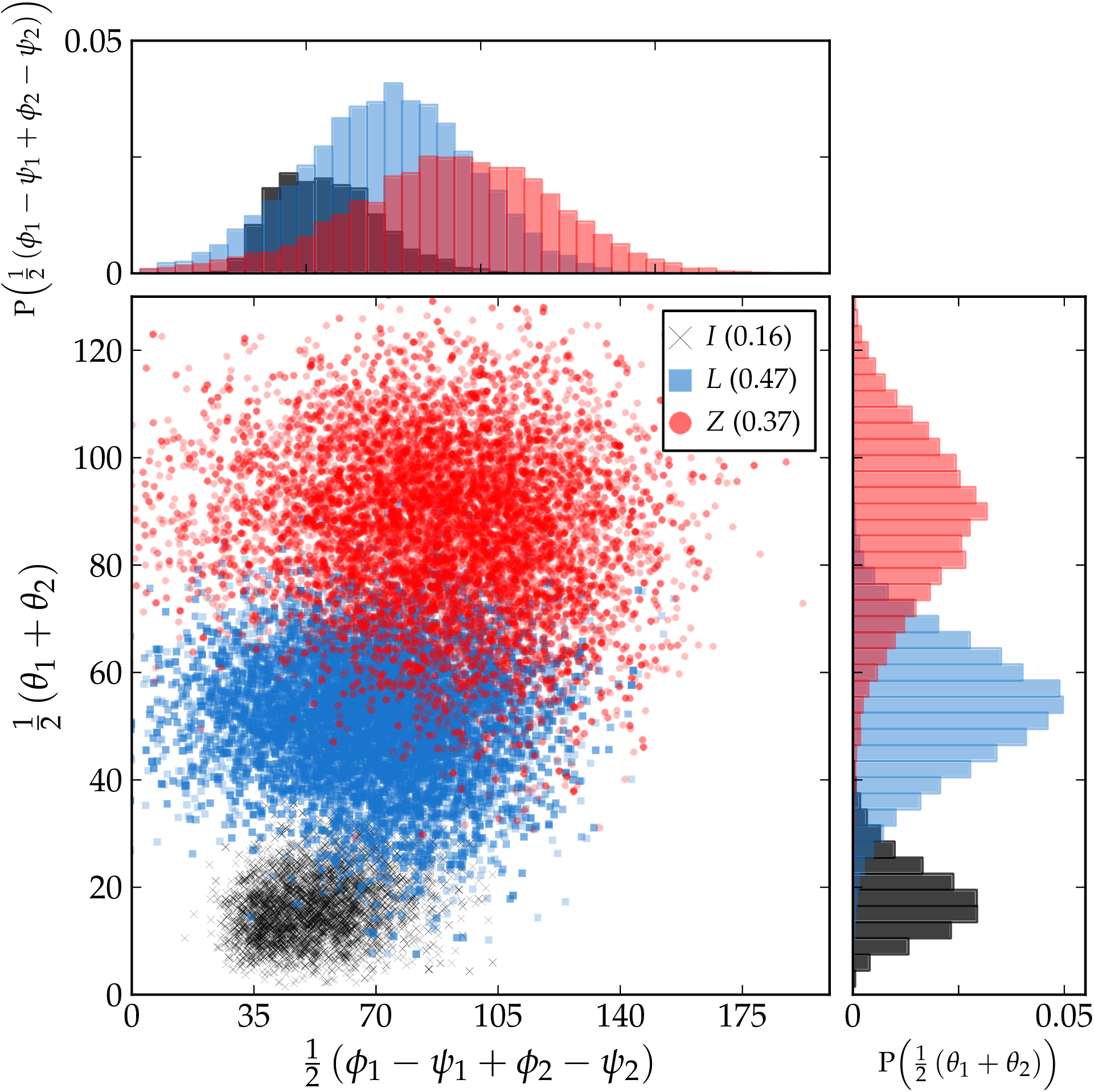}
	\caption{A scatter plot of the average bend angle versus the average twist at the bulge for different classes of the Z-tile configurations. ``I'' refers to Z-tile configurations in which both bulged regions are negligibly bent, ``L'' refers to those configurations in which one region is bent, while the other region is straight, and ``Z'' refers to those configurations in which both regions are bent at the bulge. Top panel: distributions of the average twist angles for I, L, and Z configurations. Right panel: distributions of the average bend angle of the Z-tile for I, L, and Z configurations.}
	\label{ztile_figs}
	\end{center}
	\end{figure}

As mentioned in the Section~\ref{sec:models}, the Z-tile, illustrated in Fig.~\ref{ztile}(a), is a symmetric DNA building block containing two bulge regions, three short duplex regions, and two sticky regions at the terminal ends of the tile that are complementary to the bulges. The bulge size can be exploited to control the angle in between the duplex regions. Under certain conditions, many identical copies of the Z-tile may link together where the sticky regions of one tile hybridize with the loops from another tile to form T-junctions.\cite{maoztile}  An example of a 1D ribbon built using oxDNA and containing 20 Z-tiles is illustrated in Fig.~\ref{ztile}(b). Assembled T-junctions can be seen in the structure in Fig.~\ref{ztile}(c). 

The bending and twisting angles for each bulged region in the Z-tile, as well as end-to-end distance for each duplex region flanking a bulged region, can be defined similarly to the same quantities for the bulged duplex system. A junction is defined to be unstacked if the stack opposite the bulge is broken, and stacked otherwise. We use the stacked/unstacked convention to determine if the Z-tile has 0, 1, or 2 bends. The angles and end-to-end distances are defined and illustrated in Fig.~\ref{ztile_geometry_fig} in the supplemental materials.\cite{supmat} We show the results for the angles by computing the average bend angle at the bulge in each of the two bulge regions, $\frac{1}{2}\left(\theta_1+\theta_2\right)$, and the total average twist measured at the bulges as $\frac{1}{2}\left(\phi_1-\psi_1+\phi_2-\psi_2\right)$.
 
The results for the bend and twist angles for the experimental Z-tile that we consider here, which contains 5 bases in each bulge loop, are plotted in Fig.~\ref{ztile_figs}. The plots illustrate that the bending and twisting features of the Z-tiles are similar to the same features found in bulged duplexes, namely, a Z-tile can be approximately straight (indicated by ``I'' in the figure) with a frequency of 16\%, one arm can be bent while the other arm is approximately straight and vice versa (``L'') with a frequency of 47\%, and finally both arms can be bent (``Z'') with a frequency 37\%. Predictably, the ``Z'' configurations also display the most amount of flexibility as is evidenced by the widths of the distributions for the bend and twist angles, while the bend and twisting angles for the ``I'' configurations cluster more tightly around an average value when compared with the ``Z'' configurations. Thus, when both arms are straight, the structure is quite stiff, but it can gain a significant amount of flexibility when the system breaks the stacks opposite to both of the bulge loop regions. The configurations where both arms are bent have a similar geometry to Z-tiles that have self-assembled into larger structures.

\subsection{Comparison with experiments}

FRET experiments carried out by Wo{\'z}niak \textit{et al.},\cite{wozniak2008single} yielded bend angles of \ang{32}, \ang{56}, and \ang{73}, for  $M=1,3,5$ respectively, which are broadly consistent with previous experimental results.\cite{gohlke1994kinking, dornberger1999solution, stuhmeier2000fluorescence} We see the same trend in the average bend angle predicted by oxDNA, namely, \ang{18}, \ang{20}, and \ang{60} for the same bulge sizes, but the angles are noticeably smaller. Interestingly, the experimental values are actually closer to the average oxDNA bend angle for unstacked states, namely, \ang{21}, \ang{58}, and \ang{88}, for $M=1,3,5$, respectively. This perhaps suggests that oxDNA somewhat overestimates the probability of being in the unstacked state, and could be due to overestimating the coaxial stacking strength or a neglect of local geometric factors that hinder stacking across the bulge, although we note that there are also significant uncertainties in the experimental measurements. Another possibility is that we underestimate the bend caused by base insertion, or the number of bases that can be inserted. A recent study of bulged duplexes in DNA observed an increase in flexibility of the motifs (a wider range of bend angles was observed) upon an increase in bulge loop size, findings which are broadly consistent with the predictions of oxDNA for the average bend angle distributions.\cite{shi2014structural} Interestingly, Bailor and co-workers\cite{bailor2010topology, mustoe2014coarse} have reported results for RNA where the bend angle in bulged RNA duplexes monotonically increased up to about $M=7$ and then plateaued. This is again similar to the behavior we see for the oxDNA bend angles for the unstacked configurations which rises and then plateaus at $M=8$. 

Our results are also very relevant to the self-assembly of multi-arm star tiles into polyhedra that was mentioned in the Introduction, and examples of oxDNA representations of these structures are shown in Fig.~\ref{chiral_examples}. In particular, the assembly products can depend sensitively on the bulge size with larger bulges favoring structures with larger bend angles at the vertices,\cite{he2008hierarchical} and the bulge size can also be used to influence the angles at the vertices of the final polyhedra when not fully constrained by the topology of the polyhedron, as is the case for the prism in Fig.~\ref{chiral_examples}(c). Firstly, our results show that the free-energy cost of breaking stacking at the bulge, which is necessary in the assembly of polyhedra, is relatively small for $M \ge 3$ (i.e.\ $<$ 2 $k_{\text{B}}T$), and in particular is small compared to the free-energy gain from the base-pairing associated with inter-tile assembly. Secondly, as the bulge loop gets longer, the range of bend angles available clearly increases until a point is reached where the bulge is sufficiently big that all bend angles are feasible. Simply put, larger bulges give rise to greater flexibility once the stacking at the junction is broken. For example, in the unstacked state, the free-energy cost of having a bond angle of \ang{120} (as is required for the triangles in the tetrahedron) is significantly smaller for the $M=5$ than the $M=3$ bulge (by roughly 4 $k_{\text{B}}T$), and so helps to explain why tetrahedra are not formed from 3-arm tiles with bulges of size $M=3$, but only for $M=5$. In future work we intend to study how bulge size explicitly controls the rate of closure of linear trimers to forming triangles. Similarly, in the nanoprism the lower free-energy cost of bending for larger bulges can explain how the bulge size can be used to control the angles at the vertices and hence the relative twist of the two triangular faces. A more extreme example is that 3-, 4-, 5-, and 6-arm motifs have been found to fully dimerize to form nanotubes with well-defined diameters and lengths when the bulge size is large ($M=9$), and when the tile contains hairpins in the tile arms, which further increases flexibility.\cite{qian2014self} In such structures, the bend angle is likely to be close to \ang{180}.

\section{Concluding Remarks}

We have studied the structural properties of bulged duplexes and Z-tiles using the coarse-grained oxDNA model, with the aim of understanding and characterizing the flexibility of these basic motifs that are widely used in DNA nanotechnology experiments as a means to tune the self-assembly and the equilibrium structure of the final product. We find that bulged duplexes typically adopt one of four configurations, three of which involve the duplex arms stacking at the junction and one of which does not. When loop sizes are small, the bulged duplex systems mainly prefer to be in a stacked state with some or all of the bases in the loop inserted into the helix. Base insertion has the effect of increasing the amount of twist in the duplex because of the added stacking interaction sites at the center of the system. The resultant average bending angle for small loops is modest and does not deviate too far away from the average bend angle of duplexes in the absence of bulges, which is about \ang{10} to \ang{20}.  

Once the bulge loops are about four or more nucleotides in length, configurations are increasingly found to be unstacked. In both stacked and unstacked configurations, the loop tends to be found on one side of the bulged duplex and with either the coaxial stack alone broken, or both stacks at the center of the system broken. The latter case allows the system to access greater bending angles. The model also predicts that medium sized bulges that are on the order of the persistence length of ssDNA will resist compression more strongly than longer bulges. This effect causes the system to significantly favor the unstacked configurations over the stacked configurations, with the unstacked configurations exhibiting a large static bend that is partially driven by stacking of the bulge loop with the stems. However, the range of the twisting between the duplex arms is somewhat restricted. Systems with large bulge sizes are less affected by a constrained loop, and were found to have the greatest flexibility in which the bending angle $\theta$ can assume values over a wide range, and also the relative twist between two duplex arms are free to take on nearly any value from \ang{0} to \ang{360}. We also studied the Z-tile and found similar bending and twisting features that were seen in the duplex system.  

oxDNA is a coarse-grained model that was derived to represent generic properties of DNA in a computationally efficient way. As such, it would not be expected to quantitatively reproduce all data from experiments on bulges, to which it was not explicitly parameterized. It does, however, serve to highlight the underlying physics that is of relevance to real systems. In particular, we would expect the four distinct classifications observed to be robust. Further, we have identified generic factors that drive changes in occupancy and properties of these configurations with bulge length. These factors are related to basic polymer properties and geometrical/steric constraints, and therefore will also be relevant in experiment. 

The properties that we have observed for bulged duplexes also provide insights into the use of this motif in DNA nanotechnology. In particular, the relatively small free-energy cost for unstacking for $M \ge 3$ and the greater flexibility in the unstacked state allowed by increasing bulge size can help to rationalize further the design rules for controlling the self-assembly product and final structure of polyhedra assembled from multi-arm star tiles. As is illustrated in Fig.~\ref{chiral_examples}, oxDNA is efficient enough to allow us to study these nanostructures, and understanding their self-assembly and structure will be the subjects of future work. 
  
\section*{Acknowledgments}
The authors are grateful to the Engineering and Physical Sciences Research Council for financial support, to Oxford's Advanced Research Computing and E-infrastructure South for computing support. JSS thanks Dr. Majid Mosayebi and Dr. Petr {\v{S}}ulc for helpful discussions.

\bibliography{}

\appendix

 \setcounter{figure}{0}
 \makeatletter 
 \renewcommand{\thefigure}{S\@arabic\c@figure}
 \setcounter{equation}{0}
 \renewcommand{\theequation}{S\@arabic\c@equation}
 \setcounter{table}{0}
 \renewcommand{\thetable}{S\@Roman\c@table}
  \setcounter{section}{0}
 \renewcommand{\thesection}{S\@Roman\c@section}


\section{Bend and Twist Angles}
\label{app_bend}
\begin{figure}
	\begin{center}
	\vspace{0.6 cm}
	\includegraphics[width = \columnwidth ]{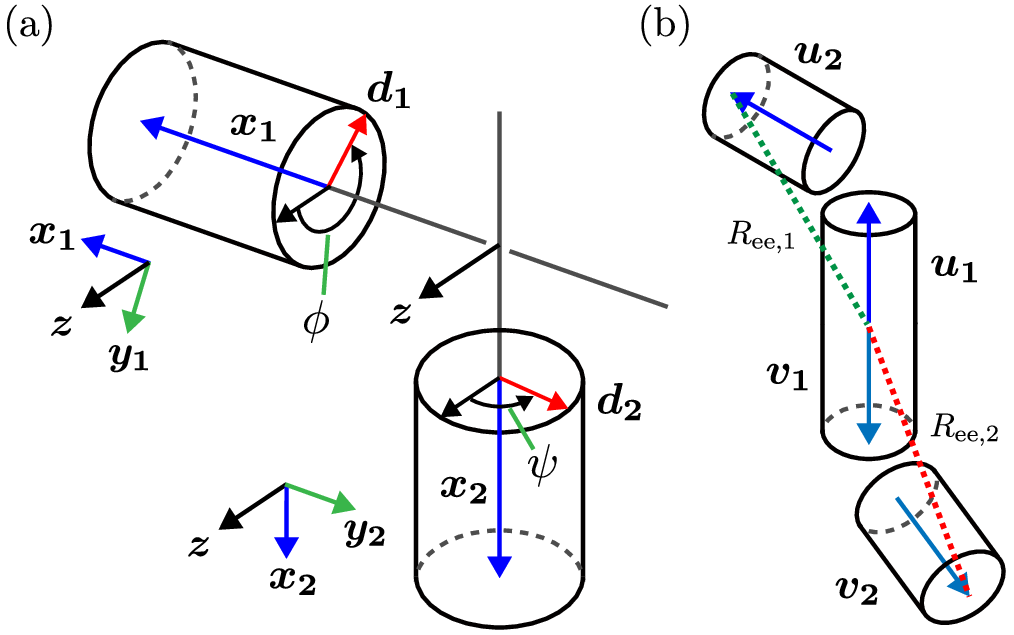}
	\caption{  In (a), the cylindrical representation of the duplex arms meeting at the bulge is illustrated, with the twist angles $\phi$ and $\psi$, the duplex unit vectors $\vect{x_{1}}$ and $\vect{x_{2}}$, and the twist vectors $\vect{d_1}$ and $\vect{d_2}$ all illustrated. The two coordinate systems are defined by crossing the normal vector $\vect{z} = \vect{x_{1}} \times \vect{x_{2}}$ with each vector $\vect{x_{1}}$ and $\vect{x_{2}}$ yielding $\vect{y_1}$ and $\vect{y_2}$, respectively. (b) Schematic illustration of a Z-tile where cylinders represent duplex regions. The end-to-end distances $R_{\text{ee,1}}$ and $R_{\text{ee,2}}$ that were used in calculating the relative free energy are also illustrated. Note that the unit vectors are not drawn to scale.}
	\label{ztile_geometry_fig}
	\end{center}
	\end{figure}

We determine whether a bulged duplex is bent into or bent away from the bulge by constructing two vectors from three defined points. The three points are taken to be the center of mass of the four bases closest to the bulge (those represented by squares and circles in Fig.~\ref{geometry} in the main text), the center of mass of the bases in the bulge, and the center of mass of the two bases at the ends of the duplex arms, whose spatial separation is $R_{\text{ee}}$. The points are referred to as A, B, and C, respectively. 
When  $\vect{BA} \cdot \vect{CA} < 0$, the system is taken to be bent away from the bulge with $0 < \theta \le 180$, and when $\vect{BA} \cdot \vect{CA} \ge 0$, the system is said to be bent into the bulge with $-180 < \theta \le 0$. For a duplex with no bulge, if $\frac{1}{2}\left(\vect{d_1}+\vect{d_2}\right) \cdot \vect{CA} \ge 0$, we take $0 < \theta \le 180$, and $-180 < \theta \le 0$ otherwise. The vectors $\vect{d_1}$ and $\vect{d_2}$ point from the center of mass of the bases at squares to the center of mass of the bases at the circles in Fig.~\ref{geometry} in the main text and are illustrated in Fig.~\ref{ztile_geometry_fig}(a), respectively.  Additionally, the relative value of the twist angles $\phi$ and $\psi$ can be determined, for example, by computing $\vect{x_1} \cdot \left( \vect{d_1} \times \vect{ \hat{z} } \right)$, where the vectors $\vect{x_1}$ and $\vect{x_2}$ were defined in Section IIE in the main text. If the quantity is greater than zero,  we let $\phi \rightarrow 360-\phi$. Similarly, if $\vect{x_2} \cdot \left( \vect{d_2} \times \vect{\hat{z}} \right) < 0$, we let $\psi \rightarrow 360-\psi$.  Despite these simplistic methods for determining the sign of $\theta$, $\phi$ and $\psi$, our simulations show that the results are reasonably reliable.  

\section{Effects of Fraying on Stacking Classes}
Here we explain how we take into account the effect of frayed base pairs near the bulge at the center of the duplex on our classification scheme. Even though most configurations analyzed are not frayed at the center, those that have become frayed can influence the interpretation of the results. If only the states of the stacks $i$ and $j$ are considered, our convention, as it stands, incorrectly associates some frayed configurations with stacked states when they should be counted as being unstacked (for example, the configuration in Fig.~\ref{stacking_cases}(e) in the main text would be counted as stacked). Additionally, fraying also interferes with our ability to properly to sort configurations into the classes \casea{}, \caseb{}, \casec{}, and \cased{}. To assist in properly sorting most frayed configurations, we instead consider the system to be ``stacked across the bulge'' when $i_{N-2} i_{N-1} i_{N} i_{N+1} i_{N+2}$ = 1, and the system is said to be ``stacked opposite the bulge'' whenever $j_{N-2} j_{N-1} j_{N} j_{N+1} j_{N+2}$ = 1 (these symbols are defined in the main text). By comparing with systems where fraying was prevented, we have found that this scheme accurately captures the state of the stacking at the center of the system in the presence of frayed base-pairs. Additionally, the configurations that are classified as \caseb{} are subjected to a second test in which the status of the following stacks are checked: $i_{N-4}, i_{N-3}, i_{N-2}, i_{N-1}$, $k_1,\dots, k_{M+1}$, $i_{N+1}, i_{N+2}, i_{N+3}, i_{N+4}$. This test ensures frayed configurations are properly sorted. If more than one of these stacks is broken, the configuration is reclassified as class \casec{}, otherwise it is left classified as \caseb{}. 

\section{Loop Stacking Effects}
\label{loop_stack_effects}

\begin{figure}
	\begin{center}
	\vspace{0.6 cm}
	\includegraphics[width = 225 pt ]{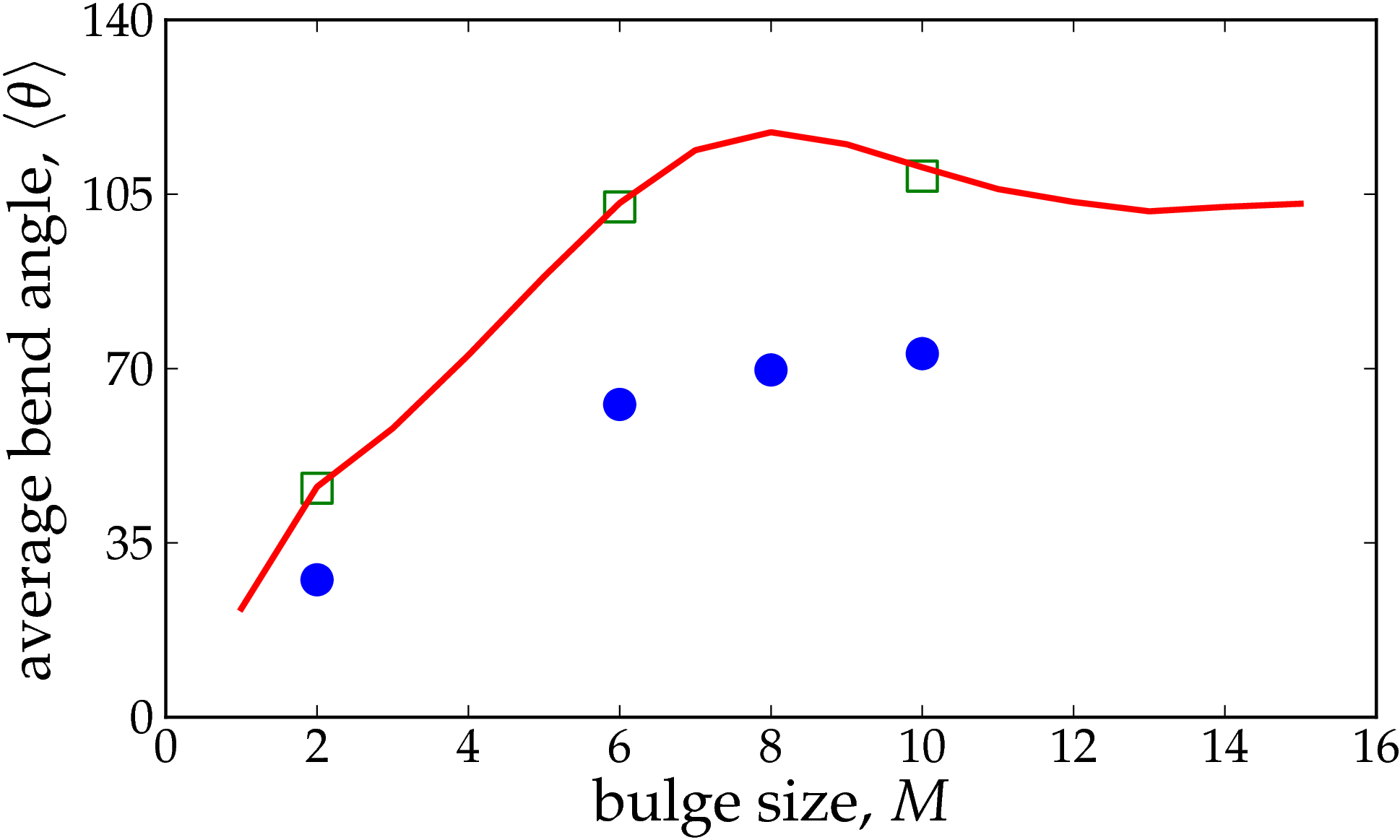}
	\caption{ Solid red line: the average bend angle for unstacked states versus bulge size, $M$. Green squares: the average bend angle versus $M$ for unstacked states when stacking between the bases in the bulge loop have been switched off. Blue circles: the average bending angle versus $M$ in which all stacking involving bases in the loop, the coaxial stack across the bulge, and the stack opposite the bulge have all been switched off. In the plot the error bar on each point is smaller than the symbol size.}
	\label{repulsive_effects}
	\end{center}
	\end{figure}

In order to determine the role that stacking between bases in the loop had on system flexibility we ran separate simulations for several bulge sizes in the three main regimes seen in the results sections: (i) small loops, where bases are often found inserted into the duplex, and where the stacked states are significantly more prevalent than unstacked states, (ii) medium sized loops, where unstacked states are more prevalent than stacked states, (iii) long loops, where there is no preference for stacked states over unstacked states, and \textit{vice versa}. To accomplish this, we ran simulations for $M=2, 6, 10$. In the simulations, we used the sequence dependent version of the model,\cite{vsulc2012sequence} which allowed us to introduce a dummy nucleotide that was used to replace the nucleotides in the loop. We were then able to switch off any stacking interaction between two dummy bases. All other interactions between a dummy nucleotide and either A, T, G or C were left unchanged. This means that in these simulations, the stem-loop stacks may still occur. The results from the simulations are plotted in Fig.~\ref{ratio_M}(a) in the main text where the free-energy difference between stacked and unstacked configurations is illustrated for several bulge sizes. When comparing the results for the system where stacking is preserved in the loop with the system where stacking is switched off in the loop, we conclude that the stacking between bases in the loop only marginally affects the probability of being unstacked at the junction, and does not significantly affect the bend angle (see Fig.~\ref{repulsive_effects}). 

The entropic repulsive effect of the stem-loop stacks on bend angle $\theta$ was also considered in the main article. In the simulations, the stacks $k_1 \cdots k_{M+1}$, $i_N$, and $j_N$ were all switched off. These results are shown in Fig.~\ref{repulsive_effects}. There is a clear reduction in the bend angle compared to that for the unstacked states in Fig.~\ref{ratio_M}(b) in the main text (which are replotted in Fig.~\ref{repulsive_effects}) and no maximum in the bend angle at intermediate values of $M$.

\section{Geometry for Z-Tile Systems}
\label{ztile_geometry}

For the Z-tile, the relative free energy was calculated as a function of two end-to-end distance order parameters, $R_{\text{ee,1}}$ and $R_{\text{ee,2}}$, which are illustrated in Fig.~\ref{ztile_geometry_fig}(b). From the Z-tile sequence used in the simulations and discussed in Section IIB of the main text, $R_{\text{ee,1}}$ was calculated by computing the distance between the center of mass of the base on the first $G$ base within the central black region (the sequence is read left-to-right) and the center of mass of the base on the final $T$ base in the terminal green region. $R_{\text{ee,2}}$ is the shortest distance between the same two bases on a second identical strand. 

The bending and twist angles are defined similarly for the Z-tile as they were for bulged duplexes. Two bending angles, $\theta_1$ and $\theta_2$ quantify the bending of each arm, respectively, and are defined as
\begin{eqnarray}
\vect{u_{1}} \cdot \vect{u_{2}} &=& - \cos \left(  \theta_1 \right), \\ 
\vect{v_{1}} \cdot \vect{v_{2}} &=& - \cos \left(  \theta_2 \right)
\end{eqnarray}
where the vectors $\vect{u_1}$, $\vect{u_2}$, $\vect{v_1}$, and $\vect{v_2}$ are illustrated in Fig.~\ref{ztile_geometry_fig}. The relative twist angle $\phi$ from one arm to the other is defined by projecting vectors $\vect{u_2}$ and $\vect{v_2}$ into the plane that is orthogonal to the vector $\vect{u_1} - \vect{v_1}$. The projections of  $\vect{u_2}$ and $\vect{v_2}$ are denoted $\vect{P_{u_2}}$ and $\vect{P_{v_2}}$, respectively. 

\section{End-to-end Probability Distribution for a Worm-Like Chain}
    \begin{figure}
	\begin{center}
	\vspace{0.6 cm}
	\includegraphics[width = 225 pt ]{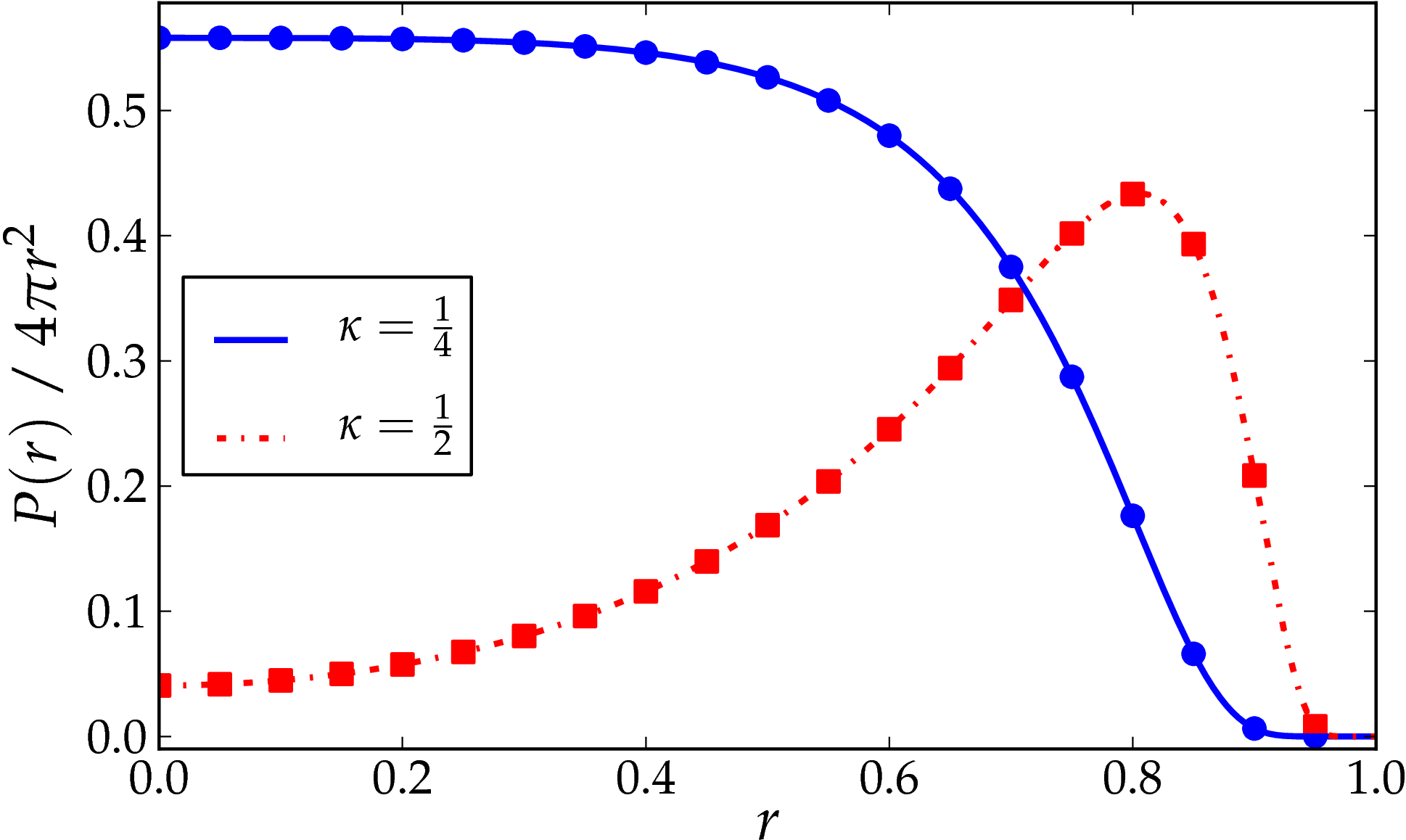}
	\caption{The interpolated probability density is plotted for worm-like chains with lengths equal to two (squares) and four (circles) times the persistence length, where $\kappa = L_p / L_c$ where $L_p$ is the persistence length and $L_c$ is the contour length of the chain. With this convention, the end-to-end separation of the chain ranges from zero to one.}
	\label{wlc_plot}
	\end{center}
	\end{figure}
In Fig.~\ref{wlc_plot}, we plot the interpolated formula for the radial end-to-end probability density of a worm-like chain as given by Eq.~21 in Ref.~\onlinecite{becker2010radial}. The parameter $r$ in Fig.~\ref{wlc_plot} is the normalized end-to-end distance of a polymer with respect to a given contour length, $L_c$. From the figure, the probability density for the chain with a contour length equal to twice its persistence length peaks near $r = 0.8$, while the probability density for the chain with contour length equal to four times its persistence length is largest for small values of $r$, with a peak at $r\approx0$. Noting that the persistence length of unstacked ssDNA in oxDNA is about 3 bases, the results for the worm-like chain help us to understand the bulged duplex structures discussed in the main text, where it was found that the longer loop benefited a lot less when having its end-to-end distance extended compared to the shorter loop. Consequently there is less incentive for longer loops to break the stacking opposite to the bulge thereby allowing the bases to spread out into space and increasing $R_{\text{ee}}$. 

\section{Additional Bend and Twist Angle Results}
	\begin{figure*}
	\begin{center}
	\vspace{0.6 cm}
	\includegraphics[width = 2\columnwidth]{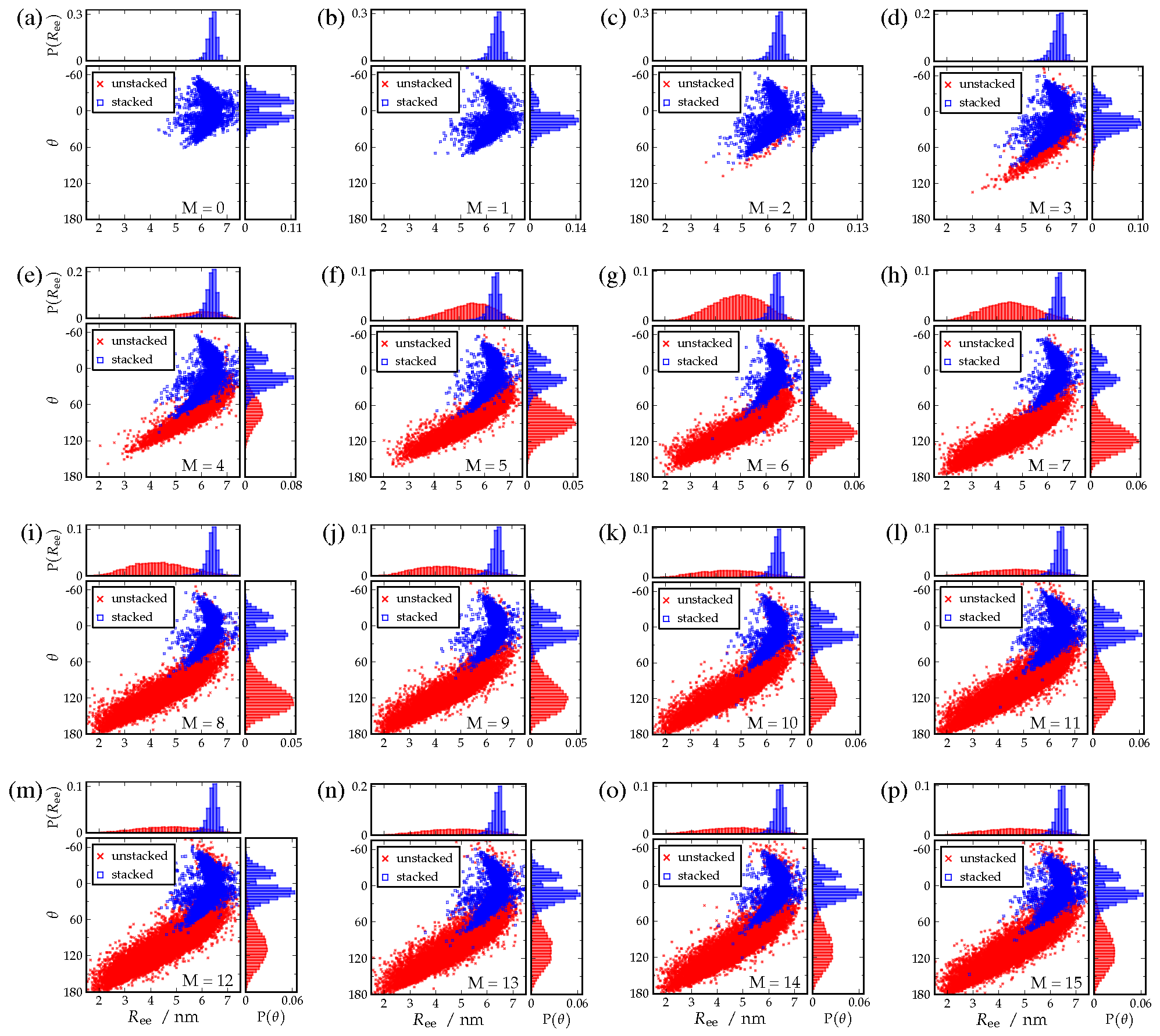}
	\caption{In each figure there are three plots that illustrate the relationship between bend angle $\theta$ and end-to-end distance $R_{\text{ee}}$: a scatter plot in ($R_{\text{ee}}$,$\theta$) for sets of configurations generated in our simulations, the probability of a state occupying a particular value of $R_{\text{ee}}$ (top panel), and the probability of a state occupying some value of the bend angle, $\theta$ (right panel). Data is plotted for a selection of bulge sizes in the range $M=0$ to $M=10$. In the scatter plots, red crosses indicate unstacked states and blue squares represent stacked states. The probability plots retain the same color scheme as the scatter plots.}
	\label{bending_plot}
	\end{center}
	\end{figure*}
	\begin{figure*}
	\begin{center}
	\vspace{0.6 cm}
	\includegraphics[width = 2\columnwidth]{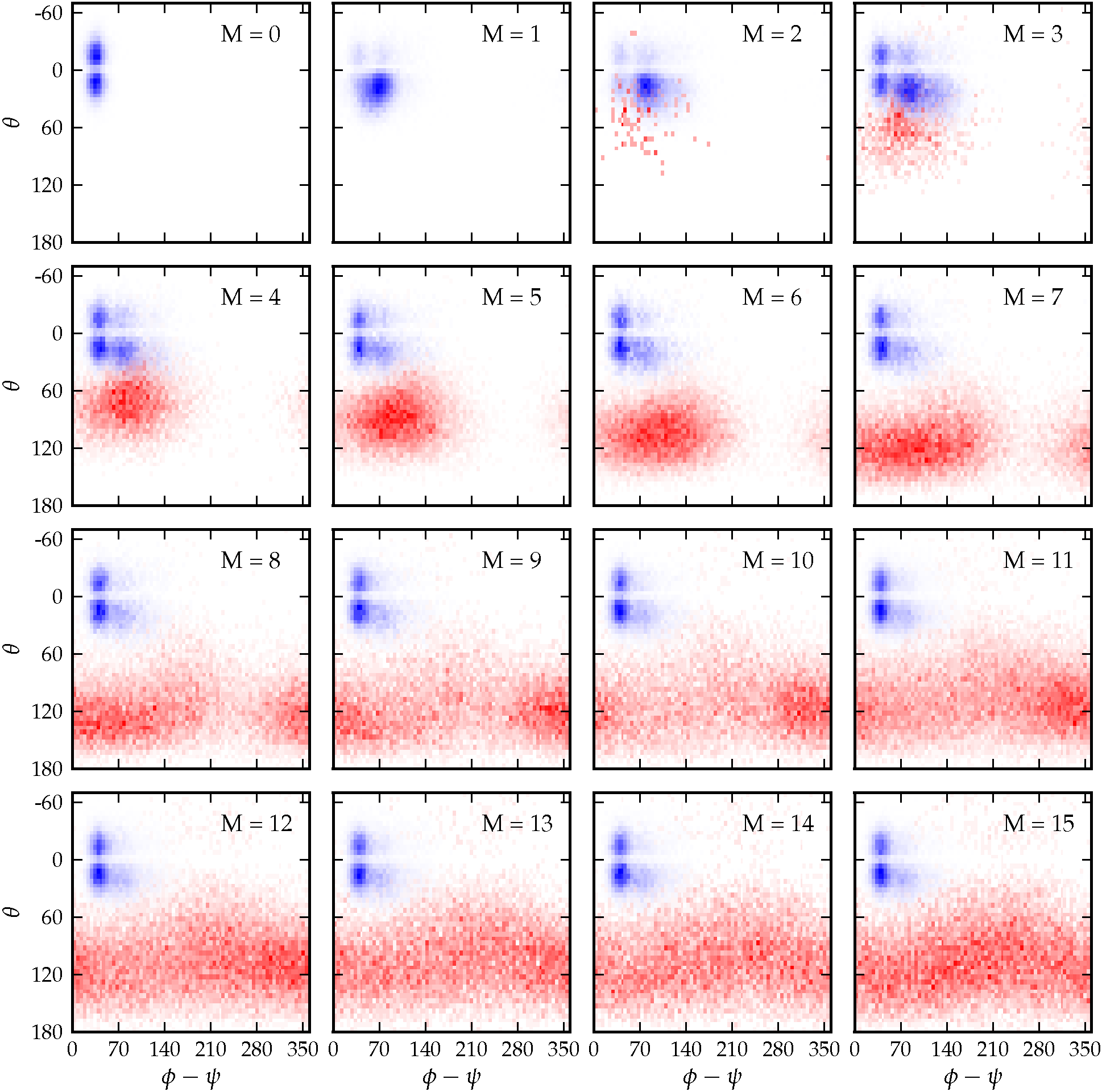}
	\caption{The probability that a configuration assumes some value of ($\theta$, $\phi$-$\psi$) is plotted for all bulge sizes studied, $M = 1-15$. Configurations were separated based on the state of the stacking between the double stranded regions, with the stacked set in blue and the unstacked set in red. The probability distributions for the stacked and unstacked sets were normalized to one separately to enhance clarity.}
	\label{twisting_plot}
	\end{center}
	\end{figure*}

In Figs.~\ref{2d_plots} and \ref{twist_plot} of the main text, we provided information on the bend and twist angle, and end-to-end distributions for a selection of bulge sizes. In Figs.~\ref{bending_plot} and ~\ref{twisting_plot} we show equivalent graphs for all bulge sizes considered (i.e. $M=1\dots15$).

\end{document}